\pdfoutput=1

\documentclass[12pt]{article}
\usepackage[utf8]{inputenc}      
\usepackage{amsmath,amssymb,mathtools}
\usepackage[T1]{fontenc}          
\usepackage{booktabs,tabularx}
\usepackage{graphicx}
\usepackage{xspace}
\usepackage{geometry}
\usepackage{todonotes}
\usepackage{listings}
\usepackage[absolute]{textpos}
\usepackage[many]{tcolorbox}
\usepackage{xparse}
\usepackage{setspace}
\usepackage{hyperref}
\usepackage[font=small,labelfont=bf,format=plain,margin=0.05\textwidth]{caption}
\usepackage{bbm}
\usepackage{tabularx}
\usepackage{subcaption}
\usepackage[normalem]{ulem}
\usepackage{soul}
\usepackage[capitalize]{cleveref}
\usepackage{slashed}
\usepackage{multirow}
\usepackage{slashed}
\usepackage[shortlabels]{enumitem}
\usepackage[sorting=none,%
  citestyle=numeric-comp,%
  bibstyle=numericstyle,%
  giveninits=true]{biblatex}

\allowdisplaybreaks

\evensidemargin \oddsidemargin
\newcommand{\GeV}{\;\text{GeV}\xspace}
\newcommand{\TeV}{\;\text{TeV}\xspace}

\newcommand{\invfb}{\ensuremath{\text{fb}^{-1}}\xspace}
\newcommand{\ma}{\ensuremath{{m_\text{a}}}\xspace}
\newcommand{\mh}{\ensuremath{{m_{\text{h}_{125}}}}\xspace}
\newcommand{\tanb}{\ensuremath{{\tan \beta}}\xspace}

\begin{document}

\thispagestyle{empty}
\def\thefootnote{\fnsymbol{footnote}}

\vspace{3em}
\begin{center}
\begin{spacing}{2.0}
{\Large\bf Search for an exotic decay of the 125 GeV Higgs boson to a pair of light pseudoscalars in the final state of two muons and two c-quarks in proton-proton collisions at $\sqrt{s} = 13\;\text{TeV}$ with CMS Open Data}
\end{spacing}

\vspace{3em}
{
Danyer Perez Adan\footnote{email: danyer.perez.adan@rwth-aachen.de}
}\\[2em]
{\sl I. Physikalisches Institut B, RWTH Aachen University,\\
Sommerfeldstra{\ss}e 16, 52074 Aachen, Germany}\\[1em]

\def\thefootnote{\arabic{footnote}}
\setcounter{page}{0}
\setcounter{footnote}{0}
\end{center}
\vspace{2ex}
\begin{abstract}
{}

A search is performed for pairs of light pseudoscalar bosons (a) produced from decays of the 125 GeV Higgs boson ($\text{h}_{125}$). The analysis is based on publicly available data collected in 2016 by the CMS experiment at the LHC in proton-proton collisions at a center-of-mass energy of 13 TeV. The amount of data analyzed corresponds to an integrated luminosity of 16.4~\ensuremath{\text{fb}^{-1}}. The analysis explores for the first time at the LHC the final state exhibiting two muons and two c-quarks, which originate from flavor-asymmetric decays of the pseudoscalar pair. The search probes the pseudoscalar boson mass interval comprised between 4 and 11 GeV, which represents a region where the light bosons exhibit a considerable Lorentz boost, and thus their decay products overlap. No significant deviation from the standard model expectation is observed.  Model-independent upper limits at 95\% confidence level are set on the product of the cross section and branching fraction for the ${\text{h}_{125} \rightarrow \text{a}\text{a} \rightarrow \mu^{-}\mu^{+} c\bar{c}}$ process relative to the standard model Higgs boson production cross section, reaching a minimum value close to $3.3 \times 10^{-4}$. The results are interpreted in the context of two Higgs doublets plus singlet models and compared to existing experimental results covering other decay channels. The exclusion limits obtained by this search improve the current constraints set by various LHC searches in scenarios where the coupling of the light boson to up-type quarks is enhanced.

\end{abstract}

\vspace{1cm}

\begin{center}
    \footnotesize
    \textit{Published in the Journal of High Energy Physics as \href{https://doi.org/10.1007/JHEP09(2025)096}{doi:10.1007/JHEP09(2025)096}}. 
\end{center}

\newpage
\tableofcontents
\newpage
\def\thefootnote{\arabic{footnote}}


\section{Introduction}
\label{sec:intro}


The discovery of a particle~\cite{ATLAS:2012yve,CMS:2012qbp} exhibiting properties similar to the Higgs boson predicted within the context of the Standard Model (SM) marked the emergence of an entirely new unexplored sector for particle physics. More than twelve years after that remarkable breakthrough, an important number of advances have been made to better understand the nature of such a particle. The mass of the predicted boson has already been measured at a remarkable precision, obtaining a value consistent with 125\GeV and an uncertainty below the 0.1\% level~\cite{CMS:2024eka,ATLAS:2023oaq}. After having been able to observe independently three of the Higgs bosonic decays using the Run 1 data with a statistical significance close to five standard deviations~\cite{CMS:2013zmy,CMS:2013fjq,CMS:2014afl}, later with larger center-of-mass energy and much more collected data during the Run 2 data-taking period, observing and accessing the direct coupling of the Higgs-like particle to the third-generation fermions became a reality~\cite{CMS:2022dwd,ATLAS:2022vkf}. Even on the relatively tiny and experimentally challenging to determine natural width of the Higgs boson, non-negligible constraints have been set already~\cite{CMS:2022ley,ATLAS:2023dnm}. Experimental studies on the spin and the parity of this new particle have shown compatibility with the SM prediction at a spectacular confidence level~\cite{CMS:2014nkk,ATLAS:2015zhl}. At the current moment, much more refined differential measurements on the various production and decay channels are also available~\cite{ATLAS:2022vkf,Belyaev:2020awq}, and none of them have evidenced a significant deviation from SM expectations.

Despite all the prominent experimental achievements mentioned above, and the success that it all represents for the SM model, now as a complete theory in its range of validity, it is notorious that the SM alone can not describe many of the experimental observations. Among those pieces of evidence, just to mention a few, there is the presence of an invisible matter (known as \textit{dark matter}) that does not interact electromagnetically, or the matter-antimatter asymmetry puzzle, which along with some theoretically unsatisfactory aspects of the model that include the absence of the gravitational interaction in its formulation, point to a beyond-SM (BSM) theory. In an attempt to address some of those unanswered questions, countless new models that partly modify the SM structure have been proposed as a potential alternative. With experimental scrutiny of the SM scalar sector having commenced not long ago, it is not a coincidence that the Higgs sector gets particularly influenced in some of the BSM models. This interest is also driven by the intrinsic versatility of scalar fields in incorporating new interactions. Some models propose
that the Higgs sector could provide a portal to dark matter~\cite{Arcadi:2019lka,Soualah:2021xbn,Cervantes:2023wti}, or may help to generate
electroweak baryogenesis of sufficient amount to explain the baryon asymmetry~\cite{Morrissey:2012db,Bodeker:2020ghk}. In the majority of the scenarios, the Higgs sector ends up being augmented, even in minimal extensions of the SM such as supersymmetric models~\cite{Martin:1997ns} or simple multi-scalar extensions~\cite{Drozd:2011aa}. Requiring an additional $SU(2)$ doublet (see e.g.~\cite{Branco:2011iw}) is a relatively simple alternative explored in some models, particularly in supersymmetry, due to the ability of this construction to provide mass simultaneously to differently-charged quarks and to cancel anomalies. In general, the structure of these two-Higgs-doublet models (2HDM) can be conceived beyond the particular case of the minimal supersymmetric model (MSSM), giving rise to richer configurations of scalar-to-fermion couplings~\cite{Curtin:2013fra}.

Although 2HDMs have received significant constraints by experimental data~\cite{Belanger:2013xza}, an extension of these models by additional scalar singlet (2HDM+S) can comfortably circumvent those restrictions if the lightest scalar mass eigenstate is identified as the SM-like 125 GeV state ($\text{h}_{125}$) and the model is assumed in the so-called decoupling limit~\cite{Gunion:2002zf}. A concrete realization of such a scalar sector can be found within the context of the next-to-minimal supersymmetric model (NMSSM)~\cite{Ellwanger:2009dp}. Within this 2HDM+S structure, there are two pseudoscalar states ($A$ and $\text{a}$), one of which (the lightest pseudoscalar, and denoted by $\text{a}$) could be very light, even lighter than the SM-like Higgs. Under these assumptions, and if the mass of the lightest pseudoscalar (\ma) satisfies the condition $\ma < \mh/2$, there could exist exotic decays of the SM-like Higgs of the form $\text{h}_{125} \rightarrow \text{a}\text{a}$, where the subsequent decay of the light boson to SM fermions takes place. This decay channel becomes relevant if the lightest pseudoscalar is weakly coupled to other particles (e.g. if a is mostly-singlet-like), in which cases the primary production of such light pseudoscalars is through Higgs exotic decays. From the experimental standpoint, the current upper bounds at 95\% confidence level (CL) on the branching fraction of the Higgs boson to undetected particles set by the ATLAS and CMS experiments are 12\% and 16\% respectively~\cite{CMS:2022dwd,ATLAS:2022vkf}, which still allows for a sufficiently large margin for those exotic decays to exist.

Numerous searches have been performed in the past to look for those exotic Higgs decays. Before the discovery of the SM-like Higgs boson, the D0 collaboration had already looked for $\text{H} \rightarrow \text{a}\text{a}$ decays in the final states containing muons and tau leptons~\cite{D0:2009aqi}. Making use of the collected Run 1 data, the CMS and ATLAS collaborations carried out searches in various mass regions~\cite{CMS:2015nay,CMS:2015twz,CMS:2017dmg,ATLAS:2015unc,ATLAS:2015rsn}, ranging from very light pseudoscalars (boosted topology) to larger masses (resolved topology) close to half the mass of the by then already found $\text{h}_{125}$ boson. At this point, the number of explored decay channels had already diversified significantly, and experimentally challenging final states such as $\tau\tau\tau\tau$ and $\gamma\gamma\gamma\gamma$ were being probed, along with other combinations like $\mu\mu b b$, that were added on top of $\mu\mu\tau\tau$ and $\mu\mu\mu\mu$ final states. The need to look for different combinations in various decay modes lies in the fact that the exact configuration of the couplings of the neutral scalar and pseudoscalar states to fermions in the general 2HDM+S can vary and it is unknown a priori~\cite{Curtin:2013fra,Craig:2012pu}. Four types (I, II, III, IV) are typically identified when requiring no flavor-changing neutral currents~\cite{Curtin:2013fra}, and within each type, the structure of the coupling to up-type quarks, down-type quarks, and charged leptons is different and dependent on the parameter \tanb \footnote{Defined as the ratio of the vacuum expectation value of the second doublet to that of the first doublet}.

Considering the above-mentioned complexity, during the Run 2 data-taking period, many more channels were incorporated in the $\text{h}_{125} \rightarrow \text{a}\text{a}$ search program of the ATLAS and CMS collaborations~\cite{CMS:2018qvj,CMS:2018zvv,CMS:2018jid,CMS:2018nsh,CMS:2019spf,CMS:2020ffa,CMS:2022xxa,CMS:2024zfv,CMS:2021pcy,ATLAS:2021ldb,ATLAS:2015unc,ATLAS:2021hbr,CMS:2018zvv,ATLAS:2018pvw,ATLAS:2020ahi,ATLAS:2018jnf}. Final states such as $bbbb$, $bb\tau\tau$, and $\gamma\gamma gg$ were also investigated in diverse production modes for the $\text{h}_{125}$ boson, depending on the experimental requirements for each of the decay channels. At the current moment, despite all the intense effort deployed by the experimental collaborations, no sign of such Higgs exotic decays has been found at a significant level.

However, despite the multiple searches, there are still regions of the 2HDM+S phase space that none of the already probed decay channels can access. This can be evidenced in some publications where the various ATLAS and CMS analyses are put together~\cite{ATL-PHYS-PUB-2025-011,Sekmen:2022vzu,Tao:2024syy} and projected into specific 2HDM+S configurations. Although the existing searches can cover ample regions in most of the model types, in some cases, like for the Type II and Type III models, none of the channels above can fully access regions of small values of \tanb and very small \ma (e.g. $\ma < 12\GeV$). This happens because, in this specific configuration, the decay of the light pseudoscalar is predominantly into c-quarks, light-quarks, and gluons~\cite{Curtin:2013fra}. In particular, for the mass region approximately (with some deviation due to non-perturbative effects~\cite{Haisch:2018kqx}) between two times the mass of the c-quark and two times the mass of the b-quark, the decay $\text{a} \rightarrow c\bar{c}$ tends to amply dominate for such low values of the \tanb parameter. Although the use of taggers based on multivariate techniques has not been uncommon~\cite{CMS:2022fyt,ATL-PHYS-PUB-2021-027,ATL-PHYS-PUB-2022-042} in the context of $\text{h}_{125} \rightarrow \text{a}\text{a}$ searches, dedicated developments for hadronic decay channels in more boosted configurations may help make the search with those more involved final states feasible.

\begin{figure}[ht]
    \centering
    \raisebox{0.3\height}{\includegraphics[width=0.49\textwidth]{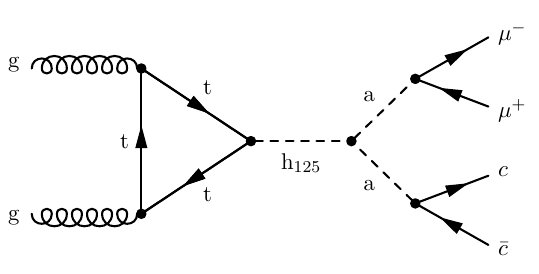}}
    \hfill
    \includegraphics[width=0.49\textwidth]{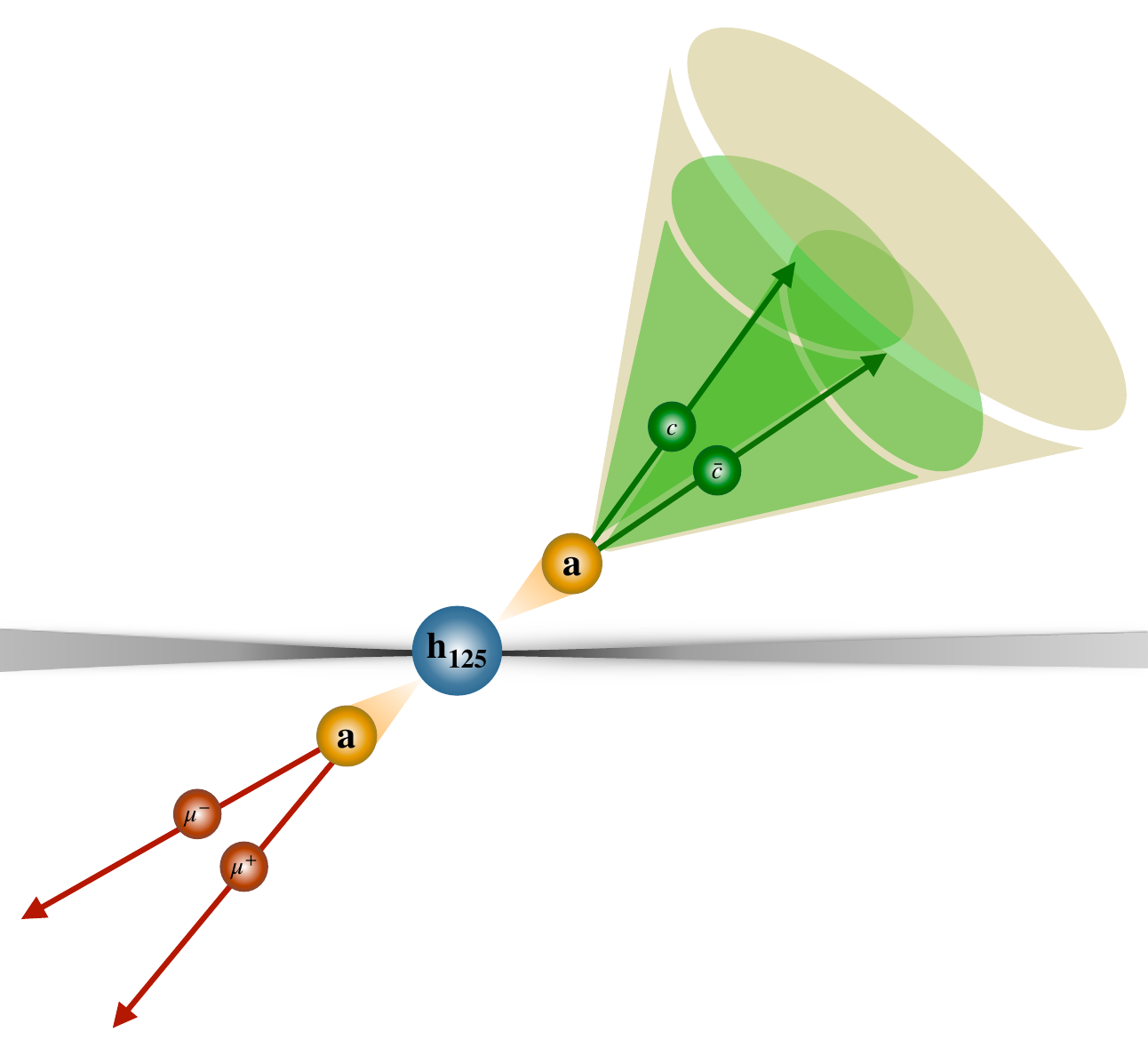}
    \caption{Feynman diagram exemplifying the production and exotic decay of the SM-like Higgs boson into a pair of pseudoscalars that subsequently decay into a $\mu^{-}\mu^{+}$ and $c\bar{c}$ pair respectively (left). Besides it, there is a schematic representation of the final state topology (right), where the effects of the boosting acquired by the pair of light bosons are illustrated.}
    \label{fig:Feynman-Diagram-and-Topology}
\end{figure}

The present study performs a dedicated search in the $\text{h}_{125} \rightarrow \text{a}\text{a} \rightarrow \mu^{-}\mu^{+} c\bar{c}$ ($\mu\mu cc$) final state using modern charm jet identification techniques~\cite{Bols:2020bkb,CMS:2021scf}, which regardless of not being optimal for $\text{a} \rightarrow c\bar{c}$ identification, may still be able to provide enough separation power to explore the uncovered phase space regions. The analysis is based on data collected in 2016 by the CMS experiment in proton-proton (pp) collisions at a center-of-mass energy of 13\TeV, and that was made publicly available by the CMS collaboration~\cite{Lassila-Perini:2021xzn,cern-open-data-portal}. The analysis has been optimized to target the main production mechanism of the SM-like Higgs boson, i.e. gluon fusion (ggF), shown in figure~\ref{fig:Feynman-Diagram-and-Topology} (left), but the contribution arising from vector boson fusion (VBF) mode is also included. The search exploits the invariant mass of the reconstructed $\text{a} \rightarrow \mu^{-}\mu^{+}$ candidate to scan for a possible excess over the expected SM background. Masses of the pseudoscalar boson between 4 and 11\GeV are probed, which represents a region where the light boson decay products are collimated and may therefore overlap, in particular, for the $\text{a} \rightarrow c\bar{c}$ leg, as illustrated in figure~\ref{fig:Feynman-Diagram-and-Topology} (right).  

This work is structured as follows. A brief description of the CMS detector is provided in Section~\ref{sec:cms_detector}. Some details about the simulated MC samples and the chosen datasets are discussed in Section~\ref{sec:mc_simulation}. Section~\ref{sec:event_selection} contains a thorough explanation of the event selection employed for this analysis, as well as some essential aspects of the event reconstruction within CMS. The modeling of the di-muon invariant mass for the signal processes here studied is explained in Section~\ref{sec:signal_modeling}, while in Section~\ref{sec:background_modeling}, the model devised for the description of background contributions is covered. Later, in Section~\ref{sec:systematics}, the treatment of the different sources of systematic uncertainties is presented. Section~\ref{sec:results} comprises the various results obtained, and finally, the work presented is summarized in Section~\ref{sec:summary}.


\section{The CMS detector}
\label{sec:cms_detector}


The data used in this analysis have been recorded with the CMS detector at the LHC in the year 2016. The distinctive component of the CMS detector~\cite{CMS:2008xjf} is a superconducting solenoid of 6 m internal diameter, which is able to supply a magnetic field of 3.8 T. The CMS detector has a cylindrical structure, symmetric around the beam pipe, and centered at the interaction point. The innermost layer is a silicon-based tracker surrounded by a scintillating crystal electromagnetic calorimeter (ECAL). After the ECAL, there is a hadron calorimeter (HCAL) followed by the outermost layer, consisting of systems designed for the detection of muons. More detailed descriptions of the CMS detector, together with a definition of the coordinate system used can be found in Ref.~\cite{CMS:2008xjf}.

Events of interest are selected using a two-tiered trigger system. The trigger system is responsible for selecting the small fraction of collision events that are relevant to the various physics activities at CMS. The CMS trigger system~\cite{CMS:2008xjf} consists of two stages: the Level-1 trigger~\cite{CMS:2000mvk}, which is entirely hardware-based and uses information from the calorimeters and muon detectors to filter events to an output rate of around 100 kHz~\cite{CMS:2020cmk}, and the software-based high-level trigger (HLT)~\cite{Sphicas:2002gg} that reduces the rate further down to around 1 kHz before data storage~\cite{CMS:2016ngn}.


\section{Selected and simulated samples}
\label{sec:mc_simulation}


The CMS experiment has released research-quality data since 2014. The CMS Open Data has already been used in several jet substructure studies and BSM searches, as documented in ref.~\cite{Lassila-Perini:2021xzn}. At the present moment, almost all reconstructed data from Run 1 is already public, along with simulated samples~\cite{cern-open-data-portal}. The first collision data at a center-of-mass energy of 13\TeV was released in 2021, while later, in 2024, about 50\% of pp data collected in 2016 at the same center-of-mass energy was released~\cite{cern-open-data-portal}. For detailed information regarding some further aspects of the CMS data preservation and open access policy see refs.~\cite{Lassila-Perini:2021xzn,cms-opendata-policy}.

This analysis is based on $pp$ collisions at a center-of-mass energy of 13 TeV collected by the CMS detector in 2016. The amount of data analyzed is equivalent to a total integrated luminosity of 16.4~\invfb. The primary datasets employed contain events recorded with muon triggers, as detailed in refs.~\cite{SingleMuon-RunG,SingleMuon-RunH}, and correspond to the last two data-taking eras (labeled \textit{G} and \textit{H}) in which the CMS detector collected $pp$ collisions in 2016.

Although the background estimation method employed in this search is fully data-driven, simulated SM background processes that contribute to the event selection were utilized to perform optimization studies and assess the overall background composition in the various analysis regions defined. As will be detailed in the next section, the most important background sources were found to be quantum chromodynamics production of multi-jet (QCD multi-jet) and Drell-Yan (DY) events. Other minor backgrounds such as top pair production ($t\bar{t}$), single-top associated production with a W boson ($tW$), and di-boson (VV, with $\text{V}=W,Z$) production were also included as part of the simulated SM background. Monte Carlo datasets simulated by the CMS collaboration were used for the above-mentioned processes - the full list of samples can be found in table~\ref{tab:mc-bkgd-samples}. The samples corresponding to VV and QCD were simulated at leading-order (LO) accuracy in QCD by CMS with the \textsc{Pythia}~\cite{Sjostrand:2014zea} (v.8.240) generator using the CP5 tune~\cite{CMS:2015wcf}. For the QCD case, the process was produced differentially in ranges of $\hat{p}_{T}$,\footnote{ The transverse momentum of each of the products, as determined in the rest frame of the colliding partons~\cite{Sjostrand:2014zea}.} as defined in \textsc{Pythia}, with an additional filter at generator level to filter events containing muons with $p_{T}>5\GeV$. The DY samples are generated at LO prediction differentially in di-lepton invariant mass and boson $p_{T}$. For an invariant mass above 10\GeV, the DY samples were generated with up to four partons in the final state using \textsc{MadGraph\_aMC@NLO}~\cite{Alwall:2014hca} (v2.6.5) with the MLM prescription~\cite{Mangano:2006rw} for matching jets from the matrix element (ME) calculation to the parton shower description. For the low-mass range in DY production, the same MC generator was used with one parton in the final state, and additionally, the phase space was divided into bins of boson $p_{T}$.
Simulated events of $t\bar{t}$ production and $tW$ process were generated at next-to-leading order (NLO) in QCD using the \textsc{Powheg}~\cite{Frixione:2007vw,Alioli:2010xd} (v2) event generator. When feasible, the predicted inclusive cross sections of the above simulated processes are corrected to match the most accurate calculations available~\cite{Li:2012wna,Czakon:2011xx,Kidonakis:2010ux,Campbell:2011bn}, which corresponds to NLO or NNLO predictions for the majority of cases. For all the above samples, \textsc{Pythia} was used to simulate parton shower, hadronization, and the underlying event~\cite{Sjostrand:2014zea}. Equally, for all simulated processes, the initial-state partons were modeled with the NNPDF 3.1 NNLO~\cite{NNPDF:2017mvq} parton distribution function (PDF), while the full CMS detector simulation was performed using \textsc{Geant4}~\cite{GEANT4:2002zbu}.

\begin{table}[ht]
\tiny
  \begin{center}
    \renewcommand{\arraystretch}{1.8}
    \begin{tabular}{|c|c|c|c|}
      \hline
      \scriptsize Process & \scriptsize Dataset name & \scriptsize Cross section [pb] & \scriptsize Filter eff. \\ 
      \hline
      \multirow{12}{*}{ \shortstack{QCD\\multi-jet}} & \multirow{1}{*}{ QCD\_Pt-15To20\_MuEnrichedPt5\_TuneCP5\_13TeV-pythia8~\cite{QCD_Pt-15To20} } & \multirow{1}{*}{ $8.488 \times 10^{8}$ } & \multirow{1}{*}{ $0.003$ } \\
      & \multirow{1}{*}{ QCD\_Pt-20To30\_MuEnrichedPt5\_TuneCP5\_13TeV-pythia8~\cite{QCD_Pt-20To30} } & \multirow{1}{*}{ $3.977 \times 10^{8}$ } & \multirow{1}{*}{ $0.006$ } \\
      & \multirow{1}{*}{ QCD\_Pt-30To50\_MuEnrichedPt5\_TuneCP5\_13TeV-pythia8~\cite{QCD_Pt-30To50} } & \multirow{1}{*}{ $1.069 \times 10^{8}$ } & \multirow{1}{*}{ $0.013$ } \\
      & \multirow{1}{*}{ QCD\_Pt-50To80\_MuEnrichedPt5\_TuneCP5\_13TeV-pythia8~\cite{QCD_Pt-50To80} } & \multirow{1}{*}{ $1.572 \times 10^{7}$ } & \multirow{1}{*}{ $0.024$ } \\
      & \multirow{1}{*}{ QCD\_Pt-80To120\_MuEnrichedPt5\_TuneCP5\_13TeV-pythia8~\cite{QCD_Pt-80To120} } & \multirow{1}{*}{ $2.343 \times 10^{6}$ } & \multirow{1}{*}{ $0.038$ } \\
      & \multirow{1}{*}{ QCD\_Pt-120To170\_MuEnrichedPt5\_TuneCP5\_13TeV-pythia8~\cite{QCD_Pt-120To170} } & \multirow{1}{*}{ $4.078 \times 10^{5}$ } & \multirow{1}{*}{ $0.052$ } \\
      & \multirow{1}{*}{ QCD\_Pt-170To300\_MuEnrichedPt5\_TuneCP5\_13TeV-pythia8~\cite{QCD_Pt-170To300} } & \multirow{1}{*}{ $1.036 \times 10^{5}$ } & \multirow{1}{*}{ $0.068$ } \\
      & \multirow{1}{*}{ QCD\_Pt-300To470\_MuEnrichedPt5\_TuneCP5\_13TeV-pythia8~\cite{QCD_Pt-300To470} } & \multirow{1}{*}{ $6.840 \times 10^{3}$ } & \multirow{1}{*}{ $0.091$ } \\
      & \multirow{1}{*}{ QCD\_Pt-470To600\_MuEnrichedPt5\_TuneCP5\_13TeV-pythia8~\cite{QCD_Pt-470To600} } & \multirow{1}{*}{ $5.526 \times 10^{2}$ } & \multirow{1}{*}{ $0.101$ } \\
      & \multirow{1}{*}{ QCD\_Pt-600To800\_MuEnrichedPt5\_TuneCP5\_13TeV-pythia8~\cite{QCD_Pt-600To800} } & \multirow{1}{*}{ $1.566 \times 10^{2}$ } & \multirow{1}{*}{ $0.116$ } \\
      & \multirow{1}{*}{ QCD\_Pt-800To1000\_MuEnrichedPt5\_TuneCP5\_13TeV-pythia8~\cite{QCD_Pt-800To1000} } & \multirow{1}{*}{ $2.628 \times 10^{1}$ } & \multirow{1}{*}{ $0.125$ } \\
      & \multirow{1}{*}{ QCD\_Pt-1000\_MuEnrichedPt5\_TuneCP5\_13TeV-pythia8~\cite{QCD_Pt-1000} } & \multirow{1}{*}{ $8.230$ } & \multirow{1}{*}{ $0.131$ } \\
      \hline
      \multirow{8}{*}{ DY } & \multirow{1}{*}{ DY1jToLL\_M-1to10\_Pt-0to70\_TuneCP5\_13TeV-madgraph-pythia8~\cite{DYJetsToLL_M-1to10_Pt-0to70} } & \multirow{1}{*}{ $1.279 \times 10^{6}$ } & \multirow{1}{*}{ -- } \\
      & \multirow{1}{*}{ DY1jToLL\_M-1to10\_Pt-70to100\_TuneCP5\_13TeV-madgraph-pythia8~\cite{DYJetsToLL_M-1to10_Pt-70to100} } & \multirow{1}{*}{ $1.345 \times 10^{1}$ } & \multirow{1}{*}{ -- } \\
      & \multirow{1}{*}{ DY1jToLL\_M-1to10\_Pt-100to200\_TuneCP5\_13TeV-madgraph-pythia8~\cite{DYJetsToLL_M-1to10_Pt-100to200} } & \multirow{1}{*}{ 4.803 } & \multirow{1}{*}{ -- } \\
      & \multirow{1}{*}{ DY1jToLL\_M-1to10\_Pt-200to400\_TuneCP5\_13TeV-madgraph-pythia8~\cite{DYJetsToLL_M-1to10_Pt-200to400} } & \multirow{1}{*}{ 0.332 } & \multirow{1}{*}{ -- } \\
      & \multirow{1}{*}{ DY1jToLL\_M-1to10\_Pt-400to600\_TuneCP5\_13TeV-madgraph-pythia8~\cite{DYJetsToLL_M-1to10_Pt-400to600} } & \multirow{1}{*}{ 0.014 } & \multirow{1}{*}{ -- } \\
      & \multirow{1}{*}{ DY1jToLL\_M-1to10\_Pt-600toInf\_TuneCP5\_13TeV-madgraph-pythia8~\cite{DYJetsToLL_M-1to10_Pt-600toInf} } & \multirow{1}{*}{ 0.002 } & \multirow{1}{*}{ -- } \\
      & \multirow{1}{*}{ DYJetsToLL\_M-10to50\_TuneCP5\_13TeV-madgraphMLM-pythia8~\cite{DYJetsToLL_M-10to50} } & \multirow{1}{*}{ $1.861 \times 10^{4}$ }& \multirow{1}{*}{ -- } \\
      & \multirow{1}{*}{ DYJetsToLL\_M-50\_TuneCP5\_13TeV-madgraphMLM-pythia8~\cite{DYJetsToLL_M-50} } & \multirow{1}{*}{ $6.077 \times 10^{3}$ } & \multirow{1}{*}{ -- } \\
      \hline
      \multirow{1}{*}{ $t\bar{t}$ } & \multirow{1}{*}{ TTTo2L2Nu\_TuneCP5\_13TeV-powheg-pythia8~\cite{TTTo2L2Nu} } & \multirow{1}{*}{ $8.731 \times 10^{1}$ } & \multirow{1}{*}{ -- } \\
      \hline
      \multirow{1}{*}{ $tW$ } & \multirow{1}{*}{ ST\_tW\_Dilept\_5f\_DR\_TuneCP5\_13TeV-amcatnlo-pythia8~\cite{ST_tW} } & \multirow{1}{*}{ 7.815 } & \multirow{1}{*}{ -- } \\
      \hline
      \multirow{3}{*}{ VV } & \multirow{1}{*}{ WW\_TuneCP5\_13TeV-pythia8~\cite{WW} } & \multirow{1}{*}{ $1.187 \times 10^{2}$ } & \multirow{1}{*}{ -- } \\
      & \multirow{1}{*}{ WZ\_TuneCP5\_13TeV-pythia8~\cite{WZ} } & \multirow{1}{*}{ $4.713 \times 10^{1}$ } & \multirow{1}{*}{ -- } \\
      & \multirow{1}{*}{ ZZ\_TuneCP5\_13TeV-pythia8~\cite{ZZ} } & \multirow{1}{*}{ $1.652 \times 10^{1}$ } & \multirow{1}{*}{ -- } \\
      \hline
    \end{tabular}
    \caption{List of selected datasets used for SM background processes along with their respective cross section before any event filtering. For those samples in which an event filter at the generator level was utilized, the efficiency values of such requirements are reported in the last column. All above MC samples are provided by the CMS collaboration under ref.~\cite{cern-open-data-portal}.}
    \label{tab:mc-bkgd-samples}
  \end{center}
\end{table}

Signal MC samples\footnote{ These are not part of the CMS Open Data and were therefore generated separately using publicly available information.} for the two (ggF and VBF) main production processes of $h_{125}$, and where the SM-like Higgs decays via $\text{h}_{125} \rightarrow \text{a}\text{a} \rightarrow \mu^{-}\mu^{+} c\bar{c}$, were generated using \textsc{MadGraph\_aMC@NLO}~\cite{Alwall:2014hca} (v2.6.5) and the \textsc{UFO} model \texttt{NMSSMHET} provided in ref.~\cite{Curtin:2013fra}. MC samples were produced for several mass hypotheses in the range from 4 to 11\GeV with a step of 1\GeV.  To account for a more accurate modeling of the $p_{T}$ spectrum of the $h_{125}$, the distribution obtained from the \textsc{MadGraph\_aMC@NLO} simulation is reweighted to match higher-order predictions. For the ggF process, the \textsc{HqT} program~\cite{Bozzi:2005wk} is employed to compute the $p_{T}$ spectrum at NNLL+NLO accuracy, while for the VBF process, the \textsc{Powheg} generator is used to derive the respective $p_{T}$ distribution at NLO precision. The CMS detector geometry and conditions employed in the simulation of the signal processes were taken to be identical to those of the above-described SM backgrounds. The parton shower, hadronization, and the underlying event were simulated using \textsc{Pythia} with the embedded CP5 tune. Equally for the PDF, the same version that CMS used for the production of the samples described above was utilized. The previously alluded details, including specific CMS software (CMSSW) version, computing environments, detector conditions, and other relevant elements used to generate the signal samples, are supplied by the CMS collaboration via the CERN Open Data portal~\cite{cern-open-data-portal}.

The effects of additional pp interactions in the same or adjacent bunch crossings (pileup) are included in all simulation samples provided by CMS and were also added to the simulation of the signal processes. A reweighting procedure is implemented to match the simulated distribution of pileup interactions with the one observed in the 2016 CMS data.



\section{Event reconstruction and selection}
\label{sec:event_selection}


The complexity and structure of the released data are the same as the one used by the CMS collaboration~\cite{cms-open-data-guide}. The newly released 2016 collision data and simulation are additionally provided in the so-called \texttt{NanoAOD} format, which embeds standard ROOT~\cite{Brun:1997pa} data structures. Condensed information for particle candidates that were reconstructed using the CMS particle-flow algorithm is stored for further processing. Legacy conditions of event reconstruction and calibration are typically present in the \texttt{NanoAOD} data format of the released data, although some specific elements of the calibration of the numerous physics objects must be observed when analyzing such a complex dataset. A more detailed explanation and dedicated documentation of the usage of the CMS Open Data can be found in ref.~\cite{cms-open-data-guide}.

In the following, a summarized description of the physics object reconstruction within CMS is provided. Emphasis is made on the primary physics objects employed in this search: muons and jets. As documented in ref.~\cite{cms-open-data-guide}, the CMS \texttt{NanoAOD} data format already contains the final physics objects as obtained from the particle-flow algorithm, which conducts the fundamental reconstruction steps described below. The requirements utilized to select the physics object candidates are specific to every analysis. The details of those selection criteria when identifying muons and jets in this search are also included below for completeness.

The information provided by the different sub-detectors in CMS is gathered and sent to the next step, which proceeds with the reconstruction and identification of all the stable particles that constitute the event. Other composite-like objects such as jets, missing transverse energy, taus, and primary (secondary) vertices are built, identified, or reconstructed from individual elements.

The particle-flow algorithm~\cite{CMS:2017yfk} is the central element to reconstruct and identify individual particles in a given event. Using a combination of the global information arising from the various elements of the CMS detector (charged particle tracks from the tracking detector, energy deposits in the HCAL and ECAL, and reconstructed tracks from the muon chambers), the multiple possible particles (electrons, muons, photons, charged hadrons, or neutral hadrons) are reconstructed and identified. The reconstructed vertex with the largest value of summed $p^{2}_{T}$ is taken to be the primary interaction vertex. The main objects used in this analysis are muons and jets, which are briefly discussed in the following.

The muons are reconstructed using the information provided by both the tracker and the muon subdetectors, employing a set of dedicated algorithms that identify tracks within the tracker or the muon system, which are later propagated to find potential matches in the alternative subsystem~\cite{CMS:2018rym}. The muon momentum is obtained from the curvature of the corresponding track by selecting one from several refits to its trajectory based on fit quality and resolution considerations. Within the primary energy range of muons arising from the potential signal here examined, the momentum resolution of muons can be as low as 1\% when they are produced in the central part of the detector. In this analysis, muons must pass the ``medium'' identification criteria~\cite{CMS:2018rym}, designed for high identification efficiency and sufficient background rejection, which corresponds to an approximate 99\% efficiency for muons in simulated W and Z events. Muons are required to have $p_{T}>5\GeV$ and $|\eta|<2.4$, as well as to pass cuts on the transverse and longitudinal impact parameters of $d_{xy}<0.2\text{ cm}$ and $d_{z}<0.5\text{ cm}$ respectively. To correct for the difference between simulation and real data, dedicated corrections for muon identification and isolation (see below for an explanation of the usage of isolation in this analysis) efficiencies are applied to simulated events, following the recommendations provided in ref.~\cite{cms-open-data-guide}. These efficiencies have been measured by the CMS collaboration using $Z \rightarrow \mu^{-}\mu^{+}$ (medium-energy muons) and $J/\psi \rightarrow \mu^{-}\mu^{+}$ (low-energy muons) events~\cite{CMS:2018rym} using the tag-and-probe method. Additionally, and because this analysis relies on the invariant mass reconstruction of a pair of muons, correction factors to improve the calibration of the muon energy scale and resolution are applied.

To reconstruct jets originated by the hadronization of quarks and gluons, firstly an identification of the charged and neutral hadrons that mostly compose them is needed. Charged hadrons are formed by the remaining tracks that do not belong to a muon or electron. Using a matching of the ECAL and HCAL energy deposits together with the track momentum, their energy and momentum can be directly determined. Neutral hadrons are identified by those HCAL energy clusters that are not linked to any charged hadron trajectory, or via a combined ECAL and HCAL energy measurement that exceeds the one expected for a charged hadron energy deposit. The energy of neutral hadrons is obtained from the corresponding corrected ECAL and HCAL energy deposits. In this analysis, jets clustered using the anti-$k_{T}$ clustering algorithm~\cite{Cacciari:2008gp} with a distance parameter of 0.4 are used (AK4), employing the corresponding identification and calibration techniques deployed by the CMS collaboration, as explained in ref.~\cite{cms-open-data-guide}. Contamination from pileup and electronic noise is subtracted using the charged-hadron subtraction method~\cite{CMS:2017yfk}. In order to reject jets coming from pileup collisions, a multivariate identification algorithm is applied on relatively low-energetic ($p_{T} < 50$ \GeV) jets~\cite{CMS:2020ebo}. The energy of reconstructed jets is corrected for effects from the detector response as a function of the jet $p_{T}$ and $\eta$ following the standard procedure~\cite{CMS:2016lmd}. Similarly, to further calibrate the resolution of the energy of reconstructed jets in simulation, a smearing procedure is performed in order to match the observed resolution in real data~\cite{CMS:2016lmd}. In this search, jets must have $p_{T} > 25$ \GeV and $|\eta| < 2.4$ to be considered further in the event selection. Jets in the vicinity ($\Delta R < 0.4$) of a selected muon are removed from the analysis, where $\Delta R = \sqrt{(\Delta\eta)^2 + (\Delta\phi)^2}$, with $\Delta\eta$ and $\Delta\phi$ the distances in pseudorapidity and azimuthal angle, respectively, between the muon and the jet.

Given the final state studied here, the charm jet identification constitutes a fundamental technique to recognize the $\text{a} \rightarrow c\bar{c}$ decays. Both the identification of c jets and b jets in CMS relies on the long lifetime and the mass of the c/b hadrons, as well as on features of tracks inside the jet (often comprising charged leptons) and the characteristics of reconstructed secondary vertices~\cite{CMS:2017wtu,CMS:2021scf}. Given that the properties of c jets tend to be somewhere in a middle point between those of light-flavor and b jets, the identification of c jets requires the usage of two discriminators. The first one is optimized to distinguish c jets from light-flavour jets (C-vs-L), whereas the other is trained to distinguish c jets from b jets (C-vs-B).  In this analysis, the DeepJet algorithm~\cite{Bols:2020bkb} devised within CMS is used as the main instrument to identify c-jets. This algorithm is a multivariate discriminator based on a deep convolutional neural network architecture. More information on the performance of this classifier in the context of the Run 2 data collected by CMS can be found in ref.~\cite{CMS:2021scf}. It is common to define several ``standard'' operating points for those algorithms on which the misidentification probabilities reach a particular value, and that can be used for analysis with different needs in terms of signal purity and efficiency. For the case of the c-tagging in CMS, three working points (WPs) are defined, based on the bi-dimensional output of the two discriminating variables (C-vs-L and C-vs-B). CMS defines three working points~\cite{cms-open-data-guide} for c-tagging: loose (L), medium (M), and tight (T). The specific values of both variables that define each WP are documented in ref.~\cite{cms-open-data-guide}. Approximately, in terms of tagging and mistagging rates, the L WP represents an identification efficiency for true c-jets close to 93\%, while it then allows for a misidentification of b and light jets of 35\% and 90\% respectively~\cite{CMS:2021scf}. The T WP, on the other hand, represents an identification efficiency for true c-jets close to 34\%, while it then mistags true b and light jets with a probability of 20\% and 3\% respectively~\cite{CMS:2021scf}. These two previously mentioned WPs are the ones utilized in this search. Corrections to account for the difference in c-tagging efficiency between simulation and data when operating on these two WPs are applied following the standard procedures and making use of the c-tagging efficiency measurements laid down by the CMS collaboration~\cite{CMS:2021scf,cms-open-data-guide}.

Events are selected using a pair of single-muon triggers, both with a $p_{T}$ threshold of 24\GeV, but with slightly different muon object requirements at the online level. The muon objects at the HLT level contain stringent requirements for identification and isolation compared to the generic muon selection described above. In order to match those tighter requirements, the offline muon with the largest transverse momentum (leading muon) is required to match the trigger object. The leading muon is then required to pass the ``tight'' identification and isolation criteria~\cite{CMS:2018rym}, as well as to have a $p_{T}$ larger than 26\GeV. In signal events, the trigger efficiency for muons satisfying such $p_{T}$ requirement was found to range from 85\% to 92\%, depending on the signal mass hypotheses and the transverse momentum of the muons. 

\begin{figure}[ht]
    \centering
    \includegraphics[width=0.49\textwidth]{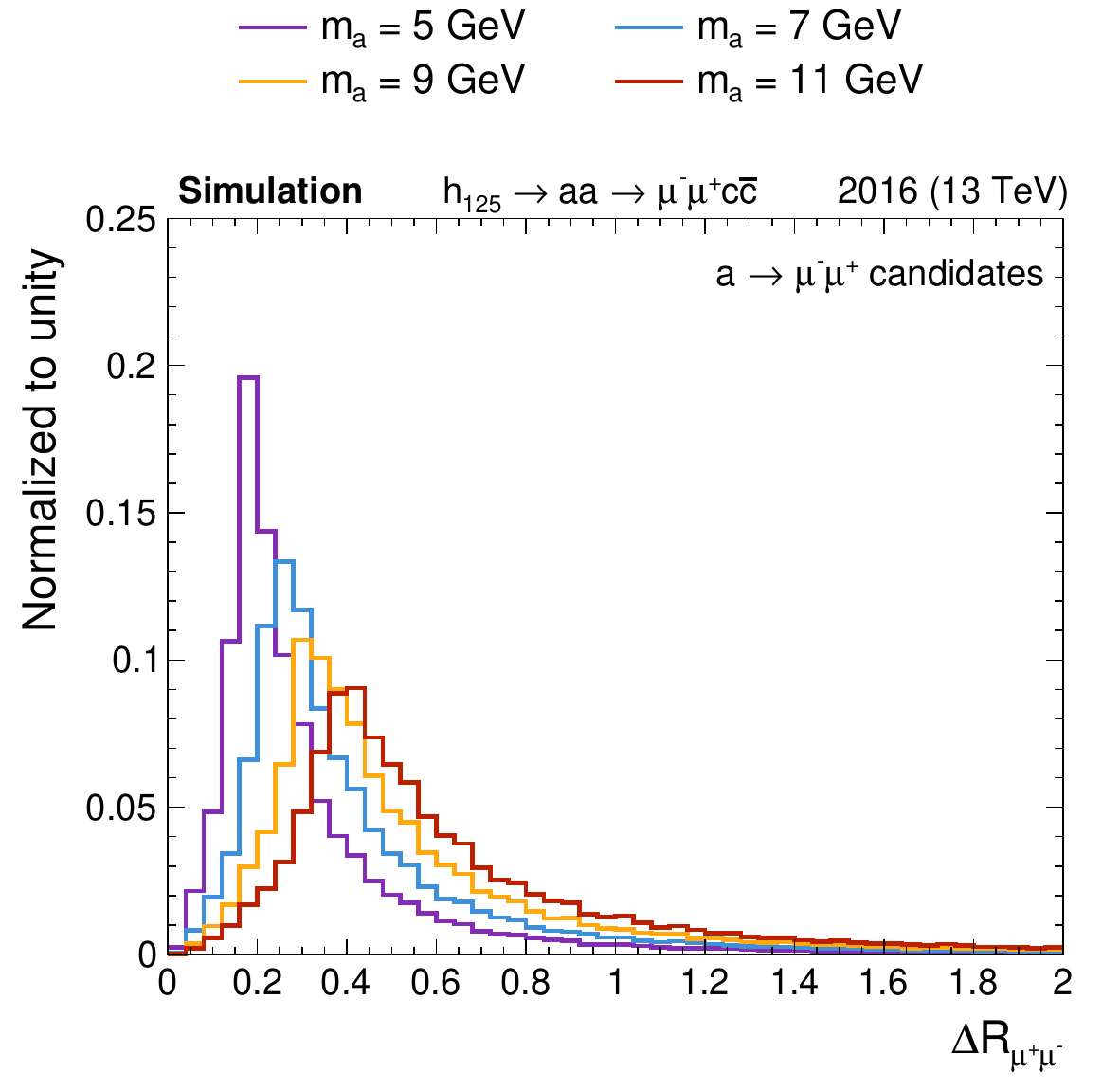}
    \hfill
    \includegraphics[width=0.49\textwidth]{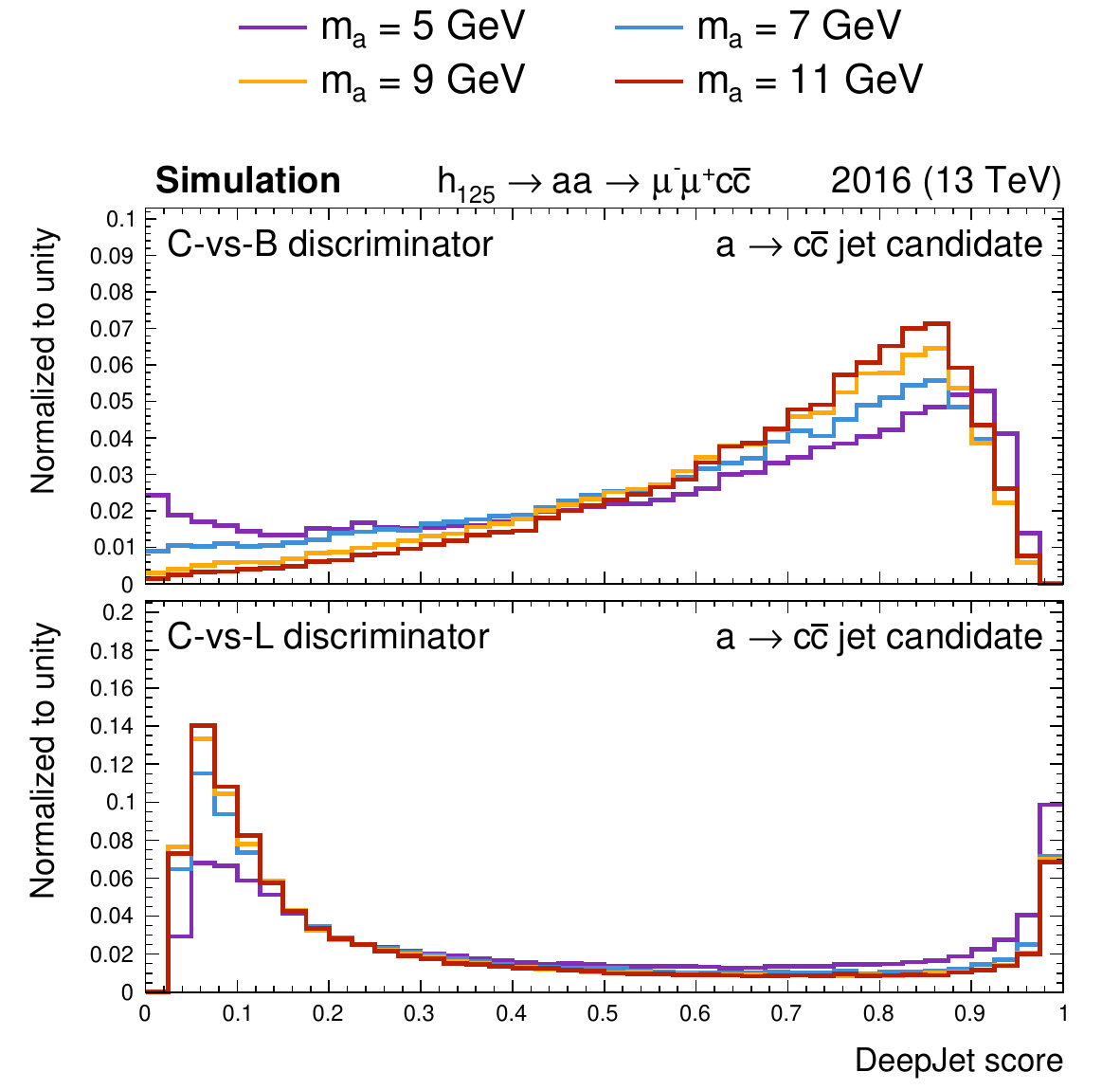}
    \caption{Relevant distributions of studies performed at generator level. On the left, the distribution of the angular separation ($\Delta R$) between the two muons that belong to the $\text{a} \rightarrow \mu^{-}\mu^{+}$ decay. On the right, the distribution of the two c-tagging discriminators (C-vs-L and C-vs-B) that are obtained for the DeepJet algorithm applied on the AK4 reconstructed jet that matches the $\text{a} \rightarrow c\bar{c}$ candidate. This matching was performed imposing that the angular distance between the direction of the light boson (truth level) and the direction of the AK4 jet (reconstructed level) is $\Delta R < 0.4$. The distributions are normalized to unity and are illustrated for four representative mass points: $\ma = 5 \GeV$ (purple line), $\ma = 7 \GeV$ (blue line), $\ma = 9 \GeV$ (yellow line), and $\ma = 11 \GeV$ (red line).}
    \label{fig:GenLevel-Studies}
\end{figure}

At the offline level, events are required to have exactly two oppositely charged muons, one of which, the leading muon, must satisfy the conditions depicted in the above paragraph. The selection continues applying further requirements on the $p_{T}$ and $\Delta R$ of the muons. The figure~\ref{fig:GenLevel-Studies} shows a study performed at generator level, where the characteristics of the $\text{a} \rightarrow \mu^{-}\mu^{+}$ and $\text{a} \rightarrow c\bar{c}$ candidates were investigated. On the left side of figure~\ref{fig:GenLevel-Studies}, one can observe the distribution of the angular separation ($\Delta R$) between the two muons originating from the $\text{a} \rightarrow \mu^{-}\mu^{+}$ leg for a few representative masses of the light boson. One can see that because of the boosting acquired by the pair of pseudoscalars, the two muons become quite collimated - the smaller the mass of the light boson, the smaller the angular separation. This characteristic of the signal was exploited to impose a requirement of $\Delta R_{\mu^{+}\mu^{-}} < 1$ and $p_{T,\mu^{+}\mu^{-}} > 40 \GeV$ on the reconstructed muon pair. From the simulated QCD multi-jet background events, it was possible to verify that, although there is a fraction of events that exhibit a similar low separation between muons, the majority of the events present a softer $p_{T,\mu^{+}\mu^{-}}$ distribution and a wider $\Delta R_{\mu^{+}\mu^{-}}$ compared to signal events. Similarly, for the DY background events, even when these tend to have a harder $p_{T,\mu^{+}\mu^{-}}$ spectrum compared to QCD multi-jet events, the di-muon angular separation in those kinds of events tends to peak for values above 3 (back-to-back configuration).

\begin{figure}[ht]
    \centering
    \includegraphics[width=0.49\textwidth]{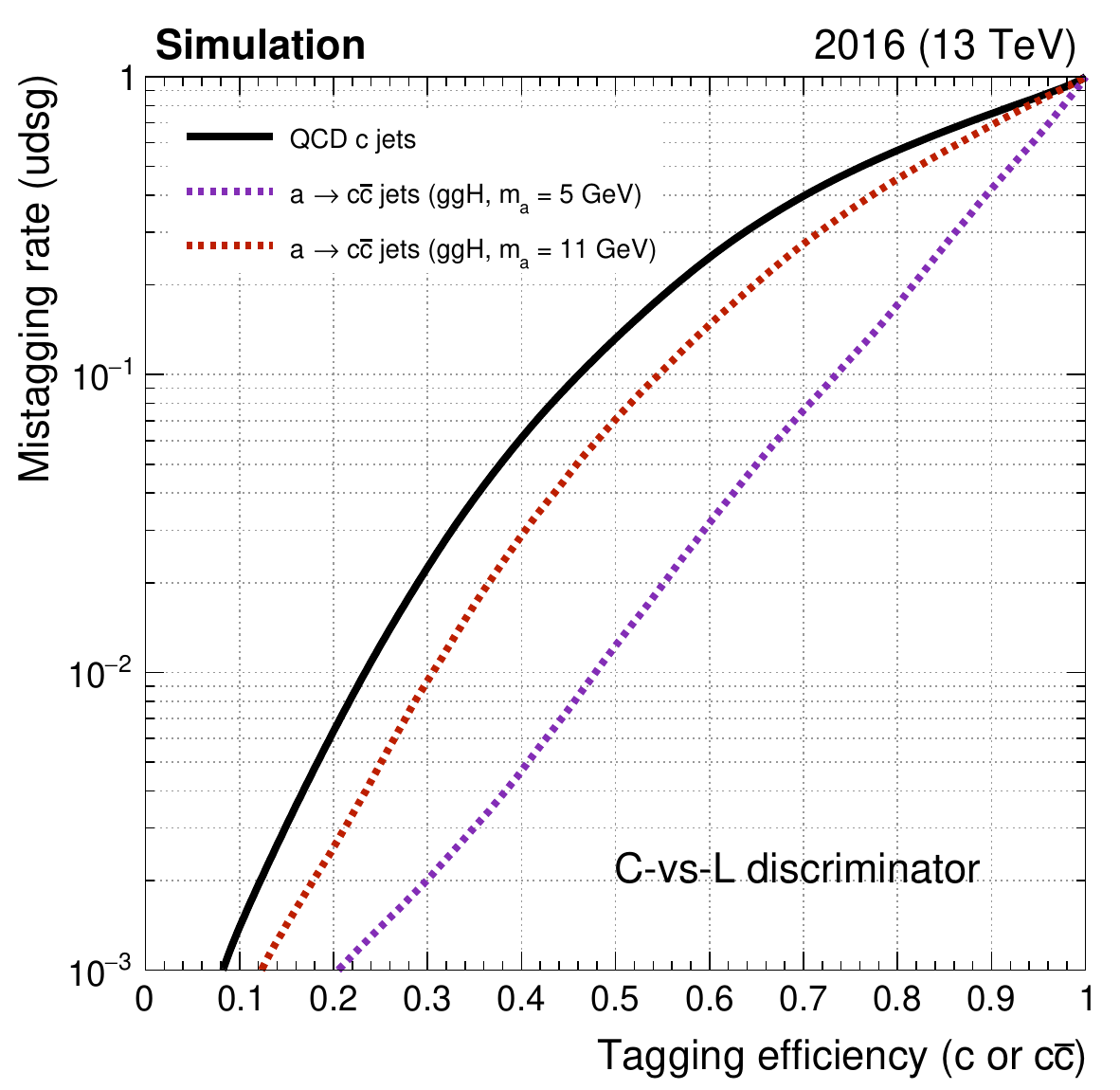}
    \hfill
    \includegraphics[width=0.49\textwidth]{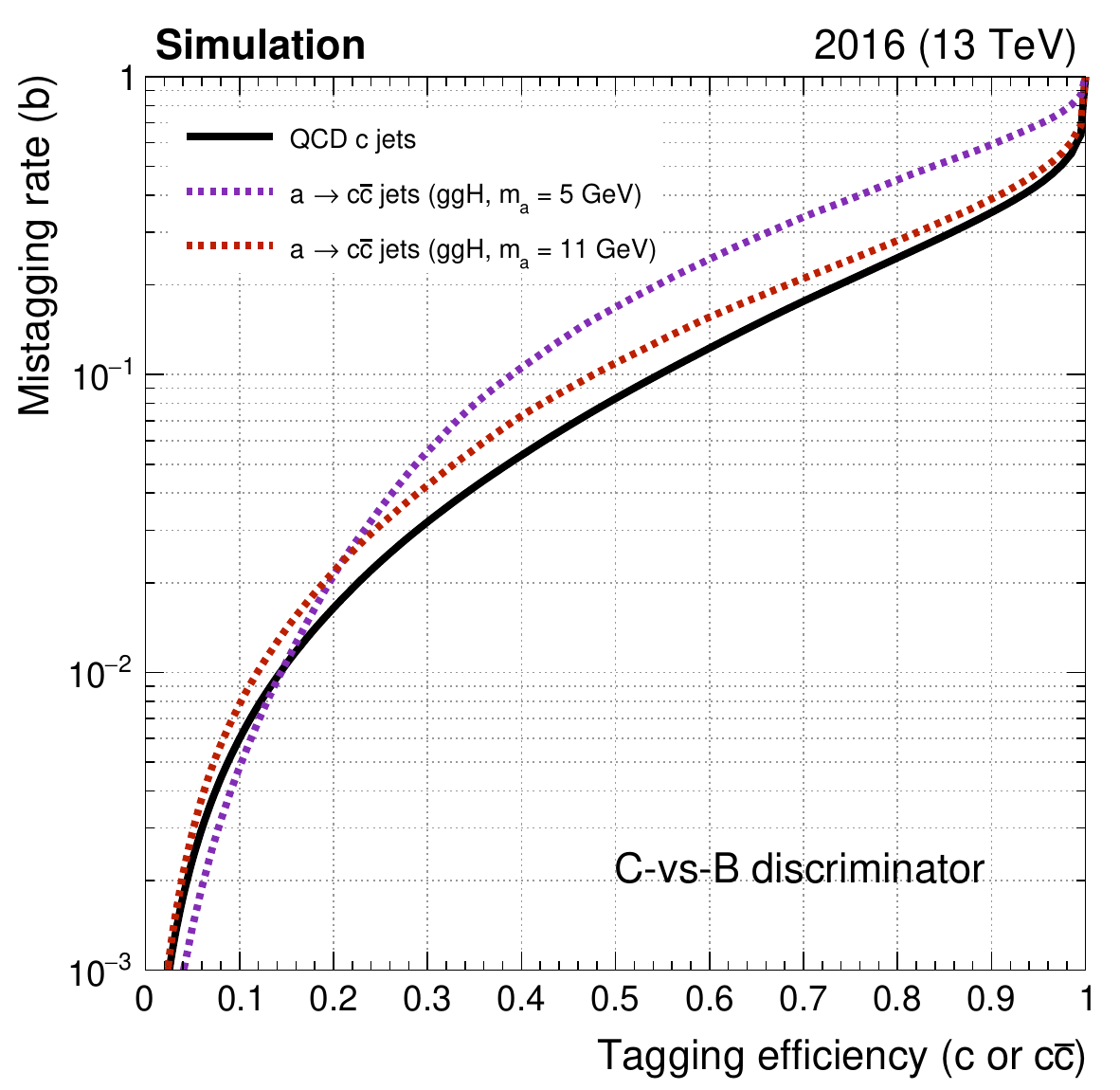}
    \caption{ROC curves exhibiting the discrimination capability of the C-vs-L (left) and C-vs-B (right) discriminators for the DeepJet algorithm using AK4 jets with similar requirements to the ones utilized in this analysis. The performance of these two discriminators when operating on signal $c\bar{c}$ jets ($\text{a} \rightarrow c\bar{c}$ jets) is compared to the performance on regular QCD c jets. For the signal process, only AK4 jets that have been matched by angular distance ($\Delta R < 0.4$) to the direction of flight of the light boson are considered true $c\bar{c}$ jets. For the estimation of the mistagging rate, regular QCD light (udsg) and b jets have been employed. Curves with dashed lines are shown for two representative mass hypotheses of the signal process: $\ma = 5 \GeV$ (purple line) and $\ma = 11 \GeV$ (red line). The black solid line has been derived using simulated jets originating from the QCD multi-jet process.}
    \label{fig:CTag-Discrimination}
\end{figure}

In the next step of the event selection, the analysis employs information about the reconstructed AK4 jets and their c-tagging classification. All the events are required to have at least one reconstructed jet that satisfies the conditions highlighted above for the case of jet objects. Analogously, for the $\text{a} \rightarrow c\bar{c}$, a dedicated simulation study was performed to evaluate the probability that a single AK4 jet can be reconstructed from the merged decay products of the light boson. The results showed that, in more than 75\% of cases, it was possible to reconstruct an AK4 jet that matches in angular distance ($\Delta R < 0.4$) the direction of the pseudoscalar. Then, given this advantageous fact, an evaluation of the output from the classification of the c-tagging DeepJet algorithm was carried out on these particular jets. The findings can be seen in figure~\ref{fig:GenLevel-Studies} (right), where one can note that the distributions of both the C-vs-L and C-vs-B discriminators when operating on signal $c\bar{c}$ jets tend to resemble the classification output for true c jets~\cite{CMS:2021scf}. To quantitatively assess the performance of these two discriminators when applied to signal $c\bar{c}$ jets, the so-called receiver operating characteristic (ROC) curve is estimated for each of them. The results are shown in figure~\ref{fig:CTag-Discrimination}, where the performance of the C-vs-L and C-vs-B discriminators is compared between signal $c\bar{c}$ jets and regular QCD c jets using the simulated samples. In the case of the signal, only $c\bar{c}$ jets that are compatible with the direction of the light boson are considered for evaluation. Two substantially different signal mass hypotheses are tested. For the computation of the mistagging rate of true non-c jets, regular QCD light and b jets have been employed. The results in figure~\ref{fig:CTag-Discrimination} indicate that the C-vs-L discriminator successfully identifies signal $c\bar{c}$ jets with more prominent discrimination against light jets than when operating on regular QCD c jets. This fact is of tremendous advantage for this analysis, given that the most predominant source of background is QCD multi-jet, and naturally, large rates of light jets are normally expected in this background process without any heavy flavor tagging requirement imposed. On the other hand, the C-vs-B discriminator when employed on signal $c\bar{c}$ jets can barely approach the performance of regular QCD c jets. Nevertheless, this does not prevent the C-vs-B discriminator from effectively distinguishing between signal $c\bar{c}$ jets and QCD b jets. The above analysis means that there is a non-negligible separation power of the DeepJet c-tagging discriminators for the case of a merged jet formed by decay products coming from the $\text{a} \rightarrow c\bar{c}$ leg. This partial discrimination power of the standard c-tagging algorithm when applied to jets emerging from the $\text{a} \rightarrow c\bar{c}$ leg will be one of the main ingredients used in this analysis to further suppress the background contributions.

In the next stage of the event selection, a requirement of $p_{T,\text{jet}_{L}} > 40 \GeV$ is imposed on the highest-$p_{T}$ (leading) jet, similarly as also imposed on the reconstructed di-muon pair. This condition allows suppression of both QCD multi-jet and DY background events that typically have a softer $p_{T}$ spectrum for jets - the latter was corroborated using the simulated QCD multi-jet and DY samples. Given that the combination of the di-muon pair and the leading jet would reconstruct the $h_{125}$ candidate, an additional cut was applied on the invariant mass of the object formed by the addition of the four-vector of those three objects. This requirement reads as $90 \GeV < m_{\mu^{+}\mu^{-}\text{jet}_{L}} < 160 \GeV$, which helps to reduce the QCD multi-jet and DY background by almost a third while keeping the majority of the signal events.

All the above-described event selection, along with an initially loose requirement on the invariant mass of the di-muon pair, namely $2\GeV <m_{\mu^{+}\mu^{-}}  < 15\GeV$, constitutes the baseline selection for this analysis.  A summary of the above-detailed requirements can be found in table~\ref{tab:selection-summary}.

\begin{table}[ht]
  \begin{center}
    \renewcommand{\arraystretch}{1.4}
    \begin{tabular}{|lc|}
      \hline
      Quantity & Baseline selection \\ 
      \hline
      $N_{\mu}$ (opposite-charge) & $=2$ \\
      $p_{T,\mu}$ & $26/5 \GeV$ leading/trailing \\
      $p_{T,\mu^{+}\mu^{-}}$ & $>40 \GeV$ \\
      $\Delta R_{\mu^{+}\mu^{-}}$ & $<1$ \\
      $N_{\text{jet}}$ & $\geq 1$ \\
      $p_{T,\text{jet}_{L}} $ & $>40 \GeV$ \\
      $m_{\mu^{+}\mu^{-}\text{jet}_{L}}$ & $90 \GeV <m_{\mu^{+}\mu^{-}\text{jet}_{L}}  < 160 \GeV$ \\
      $m_{\mu^{+}\mu^{-}}$ & $2 \GeV <m_{\mu^{+}\mu^{-}}  < 15 \GeV$ \\
      \hline
    \end{tabular}
    \caption{A summary of the main aspects of the baseline event selection for this analysis.}
    \label{tab:selection-summary}
  \end{center}
\end{table}

The number of expected and observed events that are obtained with the preliminary selection described above are reported in table~\ref{tab:event-yields-baseline}. The expected background yields are obtained using the reported cross section values in table~\ref{tab:mc-bkgd-samples} and the total integrated luminosity recorded by the combination of triggers described above, which amounts to just under 16.4~\invfb. For the signal, the reference SM cross sections for the production of the $h_{125}$ boson in ggF and VBF are used, which are calculated to be 48.58 pb and 3.79 pb respectively~\cite{deFlorian:2227475}. In addition to that, a benchmark branching fraction for the exotic decay of the Higgs to the final state here considered of $\mathcal{B}(\text{h}_{125} \rightarrow \text{a}\text{a} \rightarrow \mu^{-}\mu^{+} c\bar{c}) = 10^{-3}$ is assumed, just as a reference for the values shown in table~\ref{tab:mc-bkgd-samples}. As one can see, the majority of the background events correspond to QCD multi-jet events, with a contribution close to 70\% of the total background. As it will be detailed in section~\ref{sec:background_modeling}, this kind of event corresponds to a large degree to a low-mass resonant QCD background composed of several quarkonium states produced inside jets that can subsequently decay to a pair of muons. The next most important contribution is low-mass DY events with an extra jet, which represents more than 29\% of the total background. The contribution from top and di-boson processes is less than 1\%.

\begin{table}[ht]
  \begin{center}
    \renewcommand{\arraystretch}{1.3}
    \begin{tabular}{|cc|}
      \hline
      Process & Number of events \\ 
      \hline
      QCD multi-jet & $123833 \pm 18600$ \\
      DY & $69438 \pm 17626$ \\
      $t\bar{t}$ & $978 \pm 6$ \\
      $tW$ & $100 \pm 4$ \\
      VV & $72 \pm 3$ \\
      \hline
      $\text{h}_{125} \rightarrow \text{a}\text{a} \rightarrow \mu^{-}\mu^{+} c\bar{c}$ (ggF, $\ma = 5 \GeV$) & $154 \pm 1$ \\
      $\text{h}_{125} \rightarrow \text{a}\text{a} \rightarrow \mu^{-}\mu^{+} c\bar{c}$ (ggF, $\ma = 7 \GeV$) & $163 \pm 1$ \\
      $\text{h}_{125} \rightarrow \text{a}\text{a} \rightarrow \mu^{-}\mu^{+} c\bar{c}$ (ggF, $\ma = 9 \GeV$) & $169 \pm 1$ \\
      $\text{h}_{125} \rightarrow \text{a}\text{a} \rightarrow \mu^{-}\mu^{+} c\bar{c}$ (ggF, $\ma = 11 \GeV$) & $167 \pm 1$ \\
      $\text{h}_{125} \rightarrow \text{a}\text{a} \rightarrow \mu^{-}\mu^{+} c\bar{c}$ (VBF, $\ma = 5 \GeV$) & $7.54 \pm 0.07$ \\
      $\text{h}_{125} \rightarrow \text{a}\text{a} \rightarrow \mu^{-}\mu^{+} c\bar{c}$ (VBF, $\ma = 7 \GeV$) & $8.05 \pm 0.07$ \\
      $\text{h}_{125} \rightarrow \text{a}\text{a} \rightarrow \mu^{-}\mu^{+} c\bar{c}$ (VBF, $\ma = 9 \GeV$) & $8.28 \pm 0.07$ \\
      $\text{h}_{125} \rightarrow \text{a}\text{a} \rightarrow \mu^{-}\mu^{+} c\bar{c}$ (VBF, $\ma = 11 \GeV$) & $8.29 \pm 0.07$ \\
      \hline
      Total background & $194421 \pm 25739$ \\
      \hline
      Data & $207379$ \\
      \hline
    \end{tabular}
    \caption{Expected and observed number of events after the baseline selection in the analysis for the different background and signal processes. The expected background yields are obtained using the reported cross section values in table~\ref{tab:mc-bkgd-samples}. Typical acceptance (efficiency) values for the different background MC samples reported in table~\ref{tab:mc-bkgd-samples} range from $3 \times 10^{-8}$ to $2 \times 10^{-5}$ for QCD multi-jet samples, from $2 \times 10^{-6}$ to $3 \times 10^{-2}$ for DY samples, from $7 \times 10^{-4}$ to $8 \times 10^{-4}$ for $t\bar{t}$ and $tW$ samples, and from $7 \times 10^{-6}$ to $8 \times 10^{-5}$ for VV samples. For the signal, the SM cross sections for the ggF and VBF processes are used. Additionally, a benchmark branching fraction of $\mathcal{B}(\text{h}_{125} \rightarrow \text{a}\text{a} \rightarrow \mu^{-}\mu^{+} c\bar{c}) = 10^{-3}$ is assumed. The uncertainties reported are only associated with the statistics of the MC simulation.}
    \label{tab:event-yields-baseline}
  \end{center}
\end{table}

The final search region where the signal can be extracted is constructed by placing requirements on the c-tagging properties of the leading jet. For these events, the leading jet must pass the T c-tagging WP described before. While this requirement moderately impacts the signal efficiency, reducing it by a multiplicative factor between 0.31 and 0.36 with respect to the baseline selection and depending on the mass hypothesis, the impact on the background is much larger (reduced to a 6\% of the number of expected events in the baseline selection), thus allowing to increase the signal over background ratio (S/B) by a factor between 3.9 and 4.5 times with respect to that value after the baseline selection. Finally, and because the probed masses of the light boson range from 4 \GeV to 11 \GeV, the range of the final discriminant distribution is reduced to $m_{\mu^{+}\mu^{-}} \in [2.5 \GeV, 12 \GeV]$. As it will be discussed in section~\ref{sec:background_modeling}, this choice for the mass range facilitates an adequate coverage of the resonant structure of the QCD multi-jet background, while allowing for an extra margin to fully include signal mass hypotheses close to the edges of the above-defined interval. The region defined by all the above event requirements is henceforth denominated \textit{signal region} (SR). The total number of observed events in the SR is 12996. Unfortunately, due to the insufficient number of simulated events for the main background processes, on top of the intrinsic limitations of the simulation for QCD multi-jet events and low-mass DY events, extracting meaningful values for the expected yields of these processes in the SR turns out to be unfeasible. Relying on the simulation to estimate the full structure and composition of the background for this search is therefore not possible, hence the reason why a completely data-driven method to estimate the background in the SR is devised. For the signal, values of the acceptance and the number of expected events are reported in table~\ref{tab:event-yields-sr}.

\begin{table}[ht]
  \begin{center}
    \renewcommand{\arraystretch}{1.4}
    \begin{tabular}{|ccc|}
      \hline
      Process & Acceptance ($\mathcal{A}\times \varepsilon$) & Number of events \\ 
      \hline
      $\text{h}_{125} \rightarrow \text{a}\text{a} \rightarrow \mu^{-}\mu^{+} c\bar{c}$ (ggF, $\ma = 5 \GeV$) & $0.069 \pm 0.001$ & $55.1 \pm 0.6$ \\
      $\text{h}_{125} \rightarrow \text{a}\text{a} \rightarrow \mu^{-}\mu^{+} c\bar{c}$ (ggF, $\ma = 7 \GeV$) & $0.061 \pm 0.001$ & $48.4 \pm 0.6$ \\
      $\text{h}_{125} \rightarrow \text{a}\text{a} \rightarrow \mu^{-}\mu^{+} c\bar{c}$ (ggF, $\ma = 9 \GeV$) & $0.065 \pm 0.001$ & $51.7 \pm 0.6$ \\
      $\text{h}_{125} \rightarrow \text{a}\text{a} \rightarrow \mu^{-}\mu^{+} c\bar{c}$ (ggF, $\ma = 11 \GeV$) & $0.067 \pm 0.001$ & $53.4 \pm 0.6$ \\
      \hline
      $\text{h}_{125} \rightarrow \text{a}\text{a} \rightarrow \mu^{-}\mu^{+} c\bar{c}$ (VBF, $\ma = 5 \GeV$) & $0.042 \pm 0.001$ & $2.61 \pm 0.04$ \\
      $\text{h}_{125} \rightarrow \text{a}\text{a} \rightarrow \mu^{-}\mu^{+} c\bar{c}$ (VBF, $\ma = 7 \GeV$) & $0.037 \pm 0.001$ & $2.28 \pm 0.04$ \\
      $\text{h}_{125} \rightarrow \text{a}\text{a} \rightarrow \mu^{-}\mu^{+} c\bar{c}$ (VBF, $\ma = 9 \GeV$) & $0.038 \pm 0.001$ & $2.33 \pm 0.04$ \\
      $\text{h}_{125} \rightarrow \text{a}\text{a} \rightarrow \mu^{-}\mu^{+} c\bar{c}$ (VBF, $\ma = 11 \GeV$) & $0.039 \pm 0.001$ & $2.46 \pm 0.04$ \\
      \hline
    \end{tabular}
    \caption{Acceptance (efficiency) values and number of expected events in the SR for a few representative mass hypotheses for the two given $h_{125}$ production processes. For the calculation of the number of expected events, the SM cross sections for the ggF and VBF processes are used, and a benchmark branching fraction of $\mathcal{B}(\text{h}_{125} \rightarrow \text{a}\text{a} \rightarrow \mu^{-}\mu^{+} c\bar{c}) = 10^{-3}$ is assumed. The uncertainties reported are only associated with the statistics of the MC simulation.}
    \label{tab:event-yields-sr}
  \end{center}
\end{table}

In order to estimate the prevailing shape of the background in the SR, an additional region, where the signal is suppressed with respect to the background, is constructed. This region, denominated as \textit{control region} (CR), is defined by requiring the leading jet to fail the T requirement on the c-tagging classifier, while it is still required to pass the L WP. A simple assessment from simulation confirms that with this condition, the S/B ratio is reduced in the CR by a factor of approximately 10 with respect to the SR. Furthermore, the fact that the differential condition between the SR and the CR is solely based on the $\text{a} \rightarrow c\bar{c}$ jet candidate should be noted. This is of relevance because such a construction makes it unrelated to the $\text{a} \rightarrow \mu^{-}\mu^{+}$ leg. Specifically, the requirement is decorrelated from the $m_{\mu^{+}\mu^{-}}$ observable, thus allowing to determine the fundamental structure of the background in the SR from this CR - the latter with some caveats, as it will be discussed in section~\ref{sec:background_modeling}. The total number of observed events in the CR is 125709.


\section{Signal modeling}
\label{sec:signal_modeling}


As mentioned before, the signal is extracted by fitting the reconstructed di-muon mass distribution. For both the background and the signal, analytic probability density functions (p.d.f.s) are constructed. In order to estimate the functional form of the signal, simulated events are fit to various p.d.f.s that may integrate the different factors affecting the $m_{\mu^{+}\mu^{-}}$ distribution. It was found that, among the several p.d.f.s examined, the signal shape is well described by a double-sided Crystal Ball function~\cite{Gaiser:1985ix}, which consists of a double-sided Gaussian central core and a power-law function in each tail portion. The difference with respect to the standard non-double-sided Crystal Ball function is that it contains different parameters for both sides, left (L) and right (R), of the structure. The exact definition can be found in equation~\ref{eq:crystal-ball}, which is extracted from its implementation in the \textsc{RooFit} package~\cite{Brun:1997pa,Verkerke:2003ir}. The mathematical formulation of the double-sided Crystal Ball depends on seven parameters and reads

\begin{figure}[ht]
    \centering
    \includegraphics[width=0.49\textwidth]{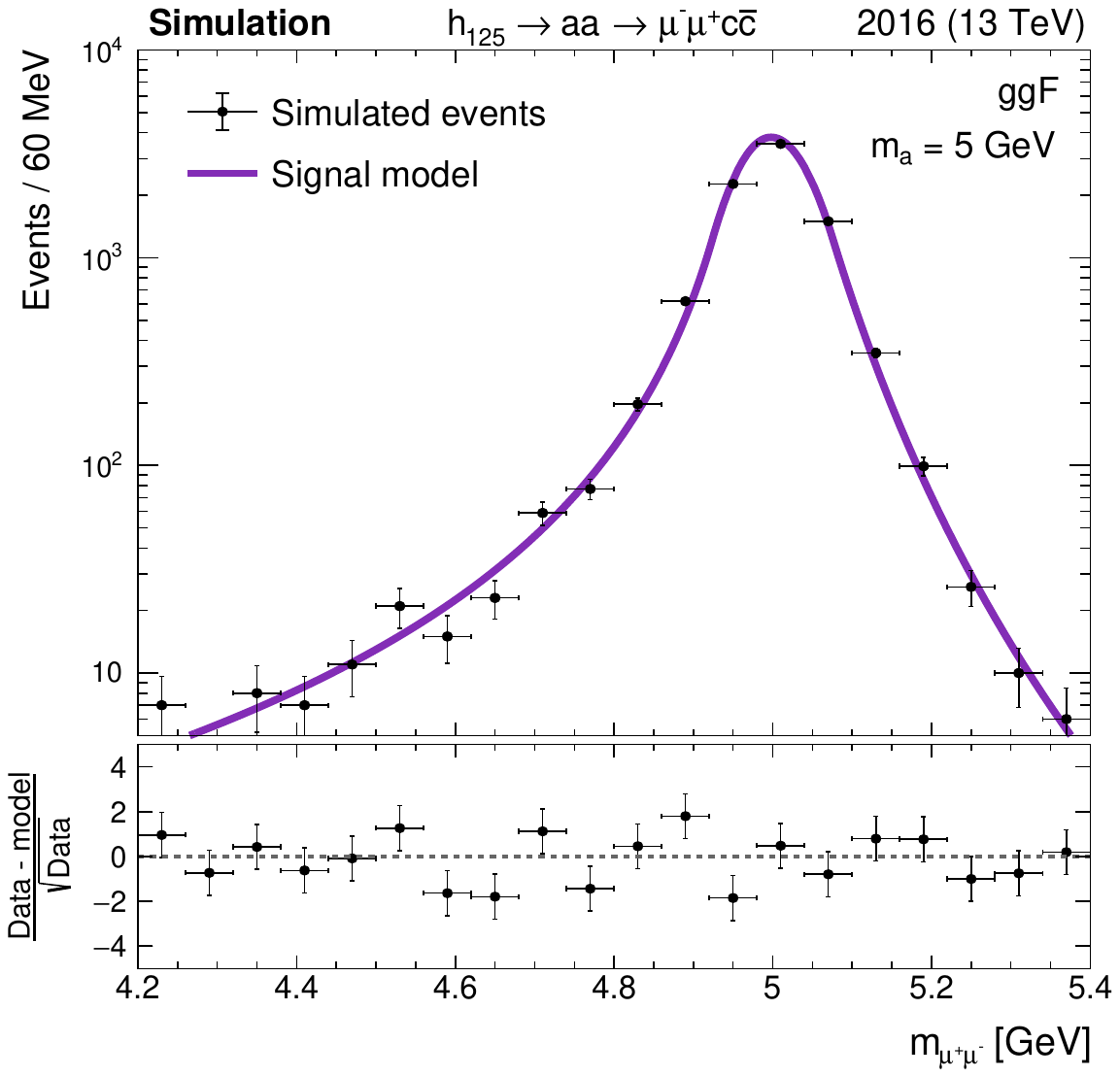}
    \hfill
    \includegraphics[width=0.49\textwidth]{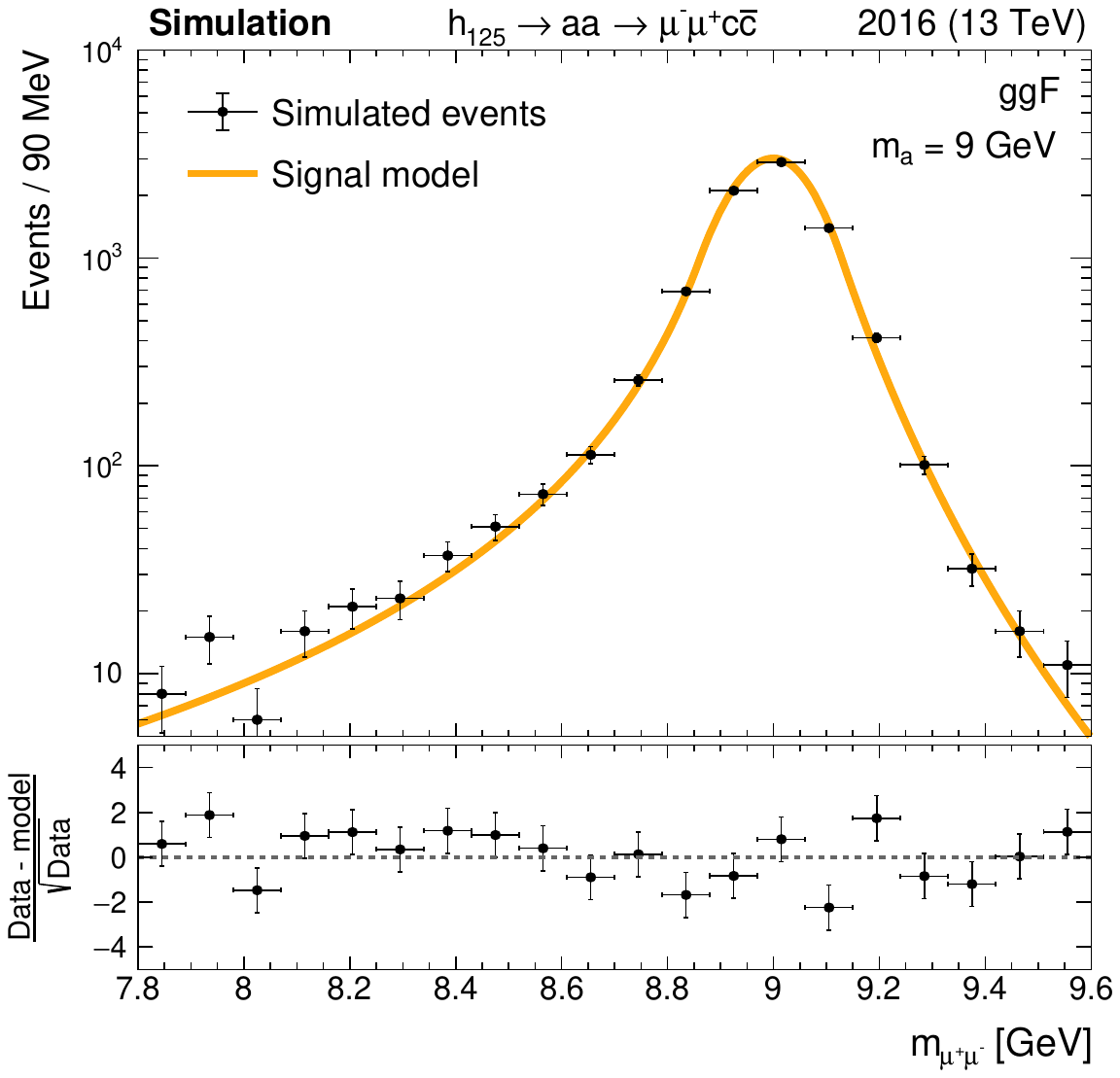}
    \caption{Graphic representation of the unbinned maximum likelihood fit performed to simulated events using the signal model underlined by equation~\ref{eq:crystal-ball} and described in the text. Two representative signal mass hypotheses, namely $\ma = 5 \GeV$ (left) and $\ma = 9 \GeV$ (right), are shown for the ggF process. The data points represent the reconstructed di-muon invariant mass distribution as obtained from simulated signal events, whereas the solid colored line represents the analytical shape of the signal after a fit was performed. The lower panel shows the standard pull obtained when comparing the fitted p.d.f. to the simulated data. The error bars on the data points include only the uncertainty related to the simulation statistics.}
    \label{fig:Signal-Model}
\end{figure}

{\small
\begin{equation}
    f(m_{\mu^{+}\mu^{-}}|m_0,\sigma_L,\sigma_R,\alpha_L,\alpha_R,n_L,n_R) = 
  \begin{cases}
    A_L \cdot (B_L - \frac{m_{\mu^{+}\mu^{-}} - m_0}{\sigma_L})^{-n_L}, & \frac{m_{\mu^{+}\mu^{-}} - m_0}{\sigma_L} < -\alpha_L \\
    \exp \left( - \frac{1}{2} \cdot \left[ \frac{m_{\mu^{+}\mu^{-}} - m_0}{\sigma_L} \right]^2 \right), & \frac{m_{\mu^{+}\mu^{-}} - m_0}{\sigma_L} \leq 0 \\
    \exp \left( - \frac{1}{2} \cdot \left[ \frac{m_{\mu^{+}\mu^{-}} - m_0}{\sigma_R} \right]^2 \right), & \frac{m_{\mu^{+}\mu^{-}} - m_0}{\sigma_R} \leq \alpha_R \\
    A_R \cdot (B_R + \frac{m_{\mu^{+}\mu^{-}} - m_0}{\sigma_R})^{-n_R}, & \mbox{otherwise},
  \end{cases}
  \label{eq:crystal-ball}
\end{equation}
}
with $A$ and $B$ normalization parameters, and defined as $A_{L/R} = (\frac{n_{L/R}}{|\alpha_{L/R}|})^{n_{L/R}} \cdot \exp(- \frac {|\alpha_{L/R}|^2}{2})$ and $B_{L/R} = \frac{n_{L/R}}{| \alpha_{L/R} |}  - | \alpha_{L/R}|$. Although, in general, the parametric dependence of equation~\ref{eq:crystal-ball} is based on seven parameters, not all of them are necessary to describe the signal shape. In fact, it was found that to reach good modeling of all considered signal scenarios, the most important feature to retain in the above equation was its differential form for the left and right sides. Therefore, it was possible to fix the following four parameters to the values $\alpha_L=1.5$, $\alpha_R=1.5$, $n_L=2.5$, and $n_R=6.5$, without affecting the quality of the goodness of fit. An example of the agreement obtained using the above model for two representative mass points and the ggF production mode can be found in figure~\ref{fig:Signal-Model}. In general, it was verified that the shape of the signal for a given mass hypothesis does not depend on the $h_{125}$ production mechanisms here probed, thus the same model was used for both types of processes.

To reproduce the corresponding signal model for intermediate mass points (see section~\ref{sec:mc_simulation}) for which a simulated sample was not generated, an interpolation of the three freely floating parameters ($m_0,\sigma_L,\sigma_R$) was performed. It was found that a linear function was sufficient to describe the dependency of these three parameters as a function of the mass hypothesis under consideration. This was validated by generating an intermediate-mass point that was not included in the above linear fit and verifying that the predicted model parameters for that point when embedded in equation~\ref{eq:crystal-ball} are able to describe the simulated data. On the other hand, for the determination of the acceptance (see table~\ref{tab:event-yields-sr}) of non-simulated signal mass points, given that the dependency of the signal efficiency in the SR is more complex, a third-degree polynomial was needed. Moreover, since the signal acceptance does depend on the $h_{125}$ production mode, the determination of the parametric form of the signal acceptance was done differentially for both ggF and VBF processes.


\section{Background modeling}
\label{sec:background_modeling}


As explained in section~\ref{sec:event_selection}, the predominant background sources are QCD multi-jet events and DY low-mass events produced in association with an extra jet. The DY background features a continuum spectrum for the $m_{\mu^{+}\mu^{-}}$ distribution with no relevant resonant structure expected. The QCD multi-jet background, on the other hand, is formed by multiple resonant components and a combinatorial continuum background originated by unrelated sources of opposite charge di-muon candidates. In the mass range studied, there are five prevalent resonances, corresponding to the quarkonium states $J/\psi(1S)$, $\Psi(2S)$, $\Upsilon(S1)$, $\Upsilon(S2)$, and $\Upsilon(S3)$ - with respective approximate masses of 3.096\GeV, 3.686\GeV, 9.460\GeV, 10.023\GeV, and 10.355\GeV, according to~\cite{ParticleDataGroup:2020ssz}. Based on this, the background model is then constructed as the sum of two exponentially decaying functions plus five resonant shapes. The choice for two exponential functions is motivated by the two kinds of continuum backgrounds that potentially arise from the combinatorial QCD multi-jet background and the low-mass DY events. For the five resonant components, the same double-sided Crystal Ball p.d.f. as for the signal model (see equation~\ref{eq:crystal-ball} and related discussion in section~\ref{sec:signal_modeling}) is taken. This results in a combined background model depending on 23 different parameters, which can be summarized in the following way:
\begin{equation}
    \begin{split}
    f(m_{\mu^{+}\mu^{-}}|\lambda_{1},\lambda_{2},\{m_{0,i},\sigma_{L,i},\sigma_{R,i}\},c_{i}) &= \sum_{i=1}^{5}c_i \cdot \mathcal{CB}_{i}(m_{0,i},\sigma_{L,i},\sigma_{R,i}) \\
                        &+ c_6\cdot e^{-\lambda_{1}m_{\mu^{+}\mu^{-}}} \\
                        &+ \left(1-\sum_{i=1}^{6}c_i \right)\cdot e^{-\lambda_{2}m_{\mu^{+}\mu^{-}}} ,                
    \end{split}
  \label{eq:background-model}
\end{equation}
where $\lambda_{1}$ and $\lambda_{2}$ are the respective exponential decay constants for the two exponential functions, $c_{i}$ represent the fraction of each component in the total p.d.f.\footnote{In this case, the p.d.f. is constructed such that it is normalized to unity, thus the fractional coefficient corresponding to the last component is derived from the rest.}, $\mathcal{CB}_{i}$ corresponds to the Crystal Ball function (equation~\ref{eq:crystal-ball}) adopted for each resonances with respective parameter set $\{m_{0,i},\sigma_{L,i},\sigma_{R,i}\}$, and where $i = [1,2,3,4,5] := [J/\psi(1S)$, $\Psi(2S)$, $\Upsilon(S1)$, $\Upsilon(S2)$, $\Upsilon(S3)]$.

\begin{figure}[ht]
    \centering
    \includegraphics[width=\textwidth]{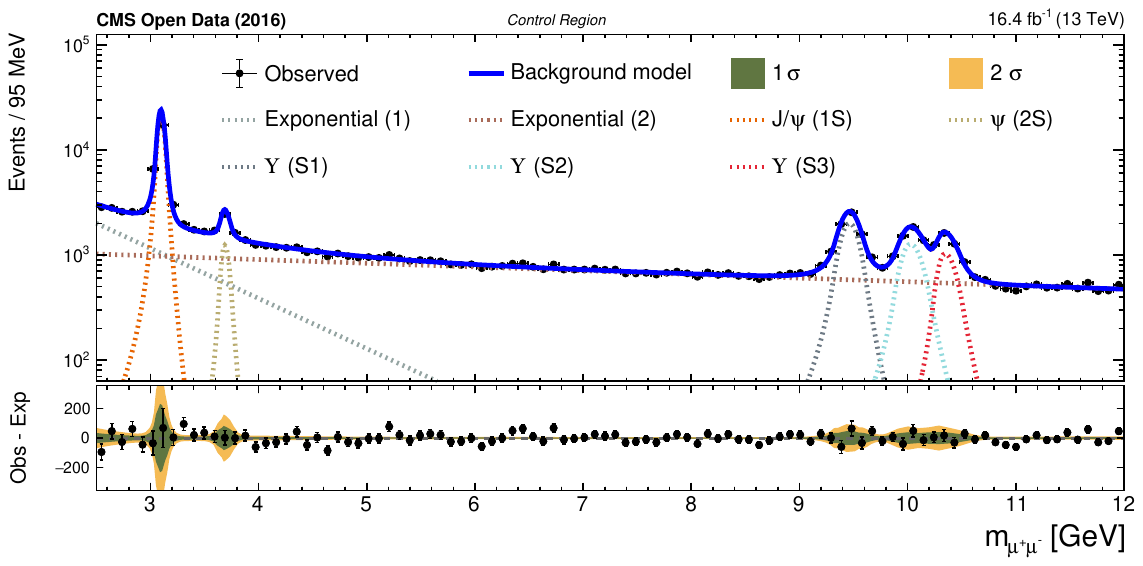}
    \caption{Illustration of the fit performed to data events selected in the CR using the background model underlined by equation~\ref{eq:background-model} and described in the text. The data points represent the reconstructed di-muon invariant mass distribution in the observed data. The blue solid line represents the full background p.d.f. after having performed a fit to CR data. The dotted lines indicate each of the different components that are embedded in the combined background p.d.f.: first exponential function (light gray), second exponential function (brown), $J/\psi(1S)$ (orange), $\Psi(2S)$ (olive green), $\Upsilon(S1)$ (dark gray), $\Upsilon(S2)$ (cyan), and $\Upsilon(S3)$ (red). The lower panel shows the difference between the combined background model conceived and the CR data. The green and yellow bands represent a visualization of the 68 and 95\% CL uncertainties in the fit. A goodness of fit test based on the \textit{saturated model} revealed a \textit{p-value} of 0.12 for the fit depicted in this figure.}
    \label{fig:Background-Model}
\end{figure}

Initially, an unbinned maximum likelihood fit is performed to the CR events, where all the parameters of equation~\ref{eq:background-model}, as well as the overall background normalization, are left freely floating. The results of this CR-only fit are depicted in figure~\ref{fig:Background-Model}, where a decomposition of the full background model into the different parts that compose it is shown. One can observe that the chosen model can describe the observed data in the CR with a high degree of accuracy, which was corroborated quantitatively (\textit{p-value} of 0.12) with a goodness of fit test based on the so-called \textit{saturated model}~\cite{CMS:2024onh}.

Now, the goal is to perform a simultaneous maximum likelihood fit including both the CR and the SR. For this, the background model in the SR is modeled with the same general structure as in the CR, and that is described by equation~\ref{eq:background-model}, but with some necessary variations. In the SR, where a signal could be present near the quarkonium mass ranges, having all the component fractions associated with the quarkonium contributions freely floating in the fit without any correlation would induce a natural bias in the model, given the similar peak structure for both the signal and the quarkonium states. To mitigate that effect, a correlation between the $c\bar{c}$ and $b\bar{b}$ bound states is devised. It is expected that the relation between the number of quarkonium events of a given type that is produced relative to the number of events of another type within the same family would not change between the two regions. Therefore, in the SR, the following restriction for the fractional components are imposed: $c^{\text{SR}}_{\Psi(2S)} = c^{\text{SR}}_{J/\psi(1S)}\cdot c^{\text{CR}}_{\Psi(2S)}/c^{\text{CR}}_{J/\psi(1S)}$, $c^{\text{SR}}_{\Upsilon(S2)} = c^{\text{SR}}_{\Upsilon(S1)}\cdot c^{\text{CR}}_{\Upsilon(S2)}/c^{\text{CR}}_{\Upsilon(S1)}$, $c^{\text{SR}}_{\Upsilon(S3)} = c^{\text{SR}}_{\Upsilon(S1)}\cdot c^{\text{CR}}_{\Upsilon(S3)}/c^{\text{CR}}_{\Upsilon(S1)}$. This means that in the combined fit between the SR and the CR the three parameters $c^{\text{SR}}_{\Psi(2S)}$, $c^{\text{SR}}_{\Upsilon(S2)}$, and $c^{\text{SR}}_{\Upsilon(S3)}$ are no longer free, but rather dependent on the remaining quarkonium fractional coefficients in both the SR and CR. That restriction allows for a correlation among the different background resonant components in the SR such that this multi-peak structure can not be biased by a single-peak signal-like appearance. The other $c_{i}$ parameters, independently in the SR and the CR, are kept freely floating in the combined fit, which includes the fractional components associated with the exponential continuous background - the latter due to the limited accuracy and size of the MC simulation, which made impossible to establish whether the background composition in the CR and SR are the same, thus a more flexible configuration was required. On the other hand, the core shape of the background resonant components is not expected to be different in the CR and SR, therefore, in order to benefit from the higher dataset size in the CR when performing the simultaneous fit, the parameter set $\{m_{0,i},\sigma_{L,i},\sigma_{R,i}\}$ is kept fully correlated (shared) between the two regions. Finally, the overall normalization of the background is inherently expected to change from the CR to SR, so it is naturally kept as an independent and unconstrained parameter in each region during the fit. The full background model described here was tested against potential residual biases using \textit{Asimov} datasets generated by injecting some amount of signal into them - the outcome of that test was successful, and thus no further addition was considered.

\section{Systematic uncertainties}
\label{sec:systematics}


Since in this analysis the estimation of the background is based on observed data, the modeling of this is not affected by imperfections in the simulation, reconstruction, or detector response. However, since most of the parameters on which the background p.d.f. model depends, including its normalization, are treated as unconstrained nuisance parameters (see section~\ref{sec:background_modeling}), given the impossibility of imposing restrictions on them based on previous knowledge (e.g. from simulation), they represent the group with the largest impact in the overall sensitivity of the analysis.

On the other hand, several systematic uncertainties affect the modeling of the signal processes here considered. The systematic uncertainties affecting the normalization of the signal are incorporated in the fit via nuisance parameters with a log-normal prior probability density function. The systematic uncertainties that affect the shape of the signal model (see section~\ref{sec:signal_modeling}), namely that change the value of the parameters that determine it, are included by adding a direct dependency into the value of the signal p.d.f. parameters, and are assigned a Gaussian prior probability density function.
 
Multiple uncertainty sources of theoretical origin impact the cross section or the acceptance of the signal processes. In the calculations of the $h_{125}$ production cross section for ggF and VBF, the unknown contributions from higher-order terms are estimated through variations in the renormalization ($\mu_{R}$) and factorization ($\mu_{F}$) scales~\cite{deFlorian:2227475}. This results in a variation of the normalization of the ggF and VBF $h_{125}$ production modes of up to 6.7\% and 0.5\% respectively. The impact of the choice of the PDF when performing such calculations was found~\cite{deFlorian:2227475} to change the predicted cross sections, and consequently the normalization of the signal, by 3.2\% and 2.1\%. The above-described uncertainties comprise only the impact that variations in $\mu_{R}$, $\mu_{F}$, and PDFs produce in the total cross section used to normalize the signal, however, these variations can also impact the signal acceptance. For the case of the PDF, its impact on signal acceptance was estimated by varying the set in the chosen NNPDF 3.1 NNLO PDF employed in the signal simulation, and it was found to be less than 1\% for both processes. In the case of the $\mu_{R}$ and $\mu_{F}$, this was done differently for ggF and VBF, though similarly varying both of them within the interval $0.5 \leq \mu_{R}$/$\mu_{F} \leq 2.0$. For the ggF, the scales were varied in the \textsc{HqT} program that was used to predict the ggF $h_{125}$ $p_{T}$ distribution, and the change that this entailed was propagated to the estimated signal acceptance (see table~\ref{tab:event-yields-sr}), yielding an overall and non-negligible impact of up to 3.9\%. For the VBF process, a similar procedure was followed, but using the respective \textsc{Powheg} prediction, which in this case represented a change of roughly 1\% in the acceptance. All the above theoretical uncertainty sources have no influence on the signal shape.

Among the experimental sources of uncertainties, the determination of the integrated luminosity can vary the total yield of both signal processes up to a value of 2.5\%~\cite{CMS:2017sdi}. The 4.8\% uncertainty associated with the measurement of the inelastic proton-proton cross section~\cite{CMS:2018mlc} impacts the pileup distribution that is used to reweight the simulated samples, which produces an overall change in the normalization of the signal processes of less than 1\%, with negligible effect on the shape of the di-muon invariant mass distribution. Other systematic uncertainties such as the muon identification, isolation, and trigger efficiency associated with the leading muon have a similar order of magnitude ($<1\%$) in the impact on the signal acceptance. On the other hand, a quite important change of between 15-17\% in the acceptance, depending on the particular signal process, was found to arise from the identification of the less energetic muon in the event. Such an effect might be related to the larger uncertainties~\cite{CMS:2018rym} obtained in the calibration measurements performed by the CMS collaboration when determining the identification efficiency of low-$p_{T}$ muons. Another relevant impact connected to muon objects, in this case affecting the shape of the signal, is the one related to the uncertainty of the muon energy scale and resolution. This implies an increase or decrease of the $p_{T}$ of muons, which in the end produces a shift in the di-muon invariant mass distribution. The three free parameters $\{m_{0},\sigma_{L},\sigma_{R}\}$ in the signal p.d.f. are affected by this uncertainty, and therefore, a functional dependency for these three parameters on a nuisance parameter assigned to the muon scale/resolution, and determined by refitting the signal for respective variations within the measured scale/resolution uncertainties, was incorporated in the fit. The remaining sources of systematic uncertainties are related to jet objects, and can thus only affect the acceptance of the signal. Among these, and as expected, the most relevant effect arises from the uncertainty on the c-tagging efficiency. The latter produces a total change of 4.7\% and 6.7\% on the signal acceptance of the ggF and VBF processes respectively. The uncertainty on the efficiency of the jet pileup identification algorithm and the uncertainty related to the jet energy resolution generate minimal effects on the normalization of less than $1\%$. The uncertainties related to the jet energy scale give rise to a change in the acceptance of 1.2\% and 0.5\% for the ggF and VBF processes respectively.

Finally, the uncertainty associated with the limited simulation statistics in the signal samples can marginally impact the values of the model parameters and the acceptance determination. The magnitude of the impact of the simulation statistics on the acceptance can be seen in table~\ref{tab:event-yields-sr}, while to account for the impact on the signal p.d.f. parameters, the post-fit uncertainties obtained were appropriately incorporated as dependencies of $\{m_{0},\sigma_{L},\sigma_{R}\}$ on three additional nuisance parameters.

\section{Results}
\label{sec:results}


As initially mentioned in section~\ref{sec:intro}, the di-muon invariant mass is scanned in a search for a potential excess of signal events, for which an unbinned maximum-likelihood fit to this distribution is performed using a combination of selected events in both the SR and CR. In this fit, both the signal normalization and the respective background normalizations (see section~\ref{sec:background_modeling}) are left freely floating. All other signal p.d.f. parameters are only allowed to vary within their respective uncertainties for a given mass hypothesis, as described in section~\ref{sec:systematics}. For the case of the background, all parameters included in equation~\ref{eq:background-model}, taking into account the particular region that they represent, are left unconstrained as well. The only constraints incorporated into the p.d.f. model describing background events in the SR are the ones above-mentioned in section~\ref{sec:background_modeling}. 

To ensure objectivity in this search, the data in the SR was kept hidden (``blinded'') during the analysis. The optimization of the event selection (see section~\ref{sec:event_selection}) for the decay channel examined in this work has been based purely on simulation. Furthermore, several diagnostic checks to investigate the robustness of the statistical model, aside from the CR-only fit described in section~\ref{sec:background_modeling}, were performed by generating pseudo-data in the SR~\cite{CMS:2024onh}.

\begin{figure}[ht]
    \centering
    \includegraphics[width=\textwidth]{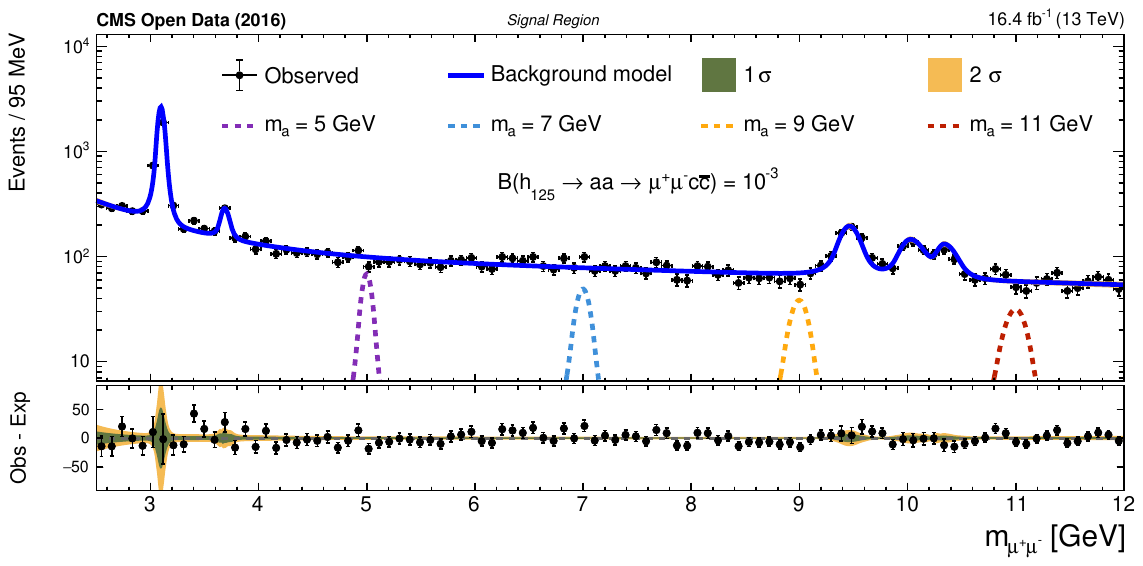}
    \caption{Invariant mass distribution of the muon pair in the SR after having performed a simultaneous background-only fit to the observed events in the SR and the CR. The black points represent the observed data, which has been binned for illustration purposes. The blue solid line represents the background prediction in the SR, while the dashed lines indicate the expected curves for four representative signal mass hypotheses: $\ma = 5\GeV$ (purple), $\ma = 7\GeV$ (cyan), $\ma = 9\GeV$ (yellow), and $\ma = 11\GeV$ (red). For this graphical representation, the signal has been normalized to the sum of the SM cross sections of the ggF and VBF processes, multiplied by a benchmark branching fraction of $\mathcal{B}(\text{h}_{125} \rightarrow \text{a}\text{a} \rightarrow \mu^{-}\mu^{+} c\bar{c}) = 10^{-3}$. The lower panel shows the difference between the observed data in the SR and the expected SM background. The green and yellow bands represent a visualization of the 68 and 95\% CL uncertainties in the fit. A goodness of fit test based on the \textit{saturated model} revealed a \textit{p-value} of 0.36 for the simultaneous fit illustrated above.}
    \label{fig:Final-Fit}
\end{figure}

In figure~\ref{fig:Final-Fit}, the di-muon invariant mass distribution in the SR is shown. The figure shows the obtained background profile after applying a simultaneous fit to the observed data in both the CR and the SR under the background-only hypothesis, as well as the expectations for the signal for a few representative mass points. As can be seen, no significant deviations from the expected SM background are observed in that distribution.

\begin{figure}[ht]
    \centering
    \includegraphics[width=0.75\textwidth]{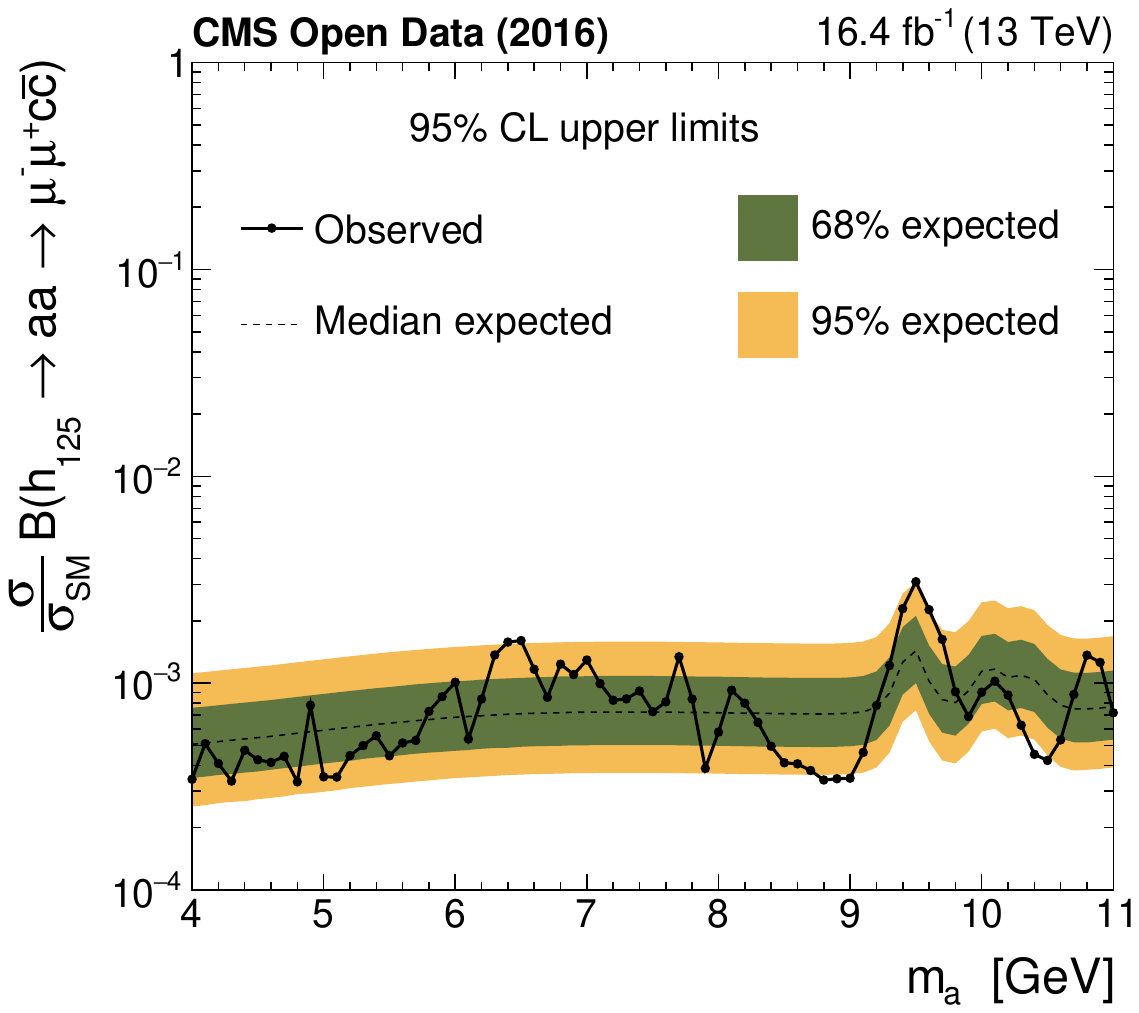}
    \caption{Observed and expected upper limits at 95\%~CL on the product of the signal cross section and branching fraction $\sigma/\sigma_{\text{SM}} \cdot \mathcal{B}(\text{h}_{125} \rightarrow \text{a}\text{a} \rightarrow \mu^{-}\mu^{+} c\bar{c})$ relative to the SM prediction. The solid and dashed lines correspond to the observed and median expected limits, respectively, while the green and yellow bands indicate the regions that contain 68\% and 95\% of the expected upper limits.}
    \label{fig:Limits-BR}
\end{figure}

Upper limits at 95\%~CL are set on the combined product of the production cross section and branching fraction relative to the SM Higgs boson production cross section, namely $\sigma/\sigma_{\text{SM}} \cdot \mathcal{B}(\text{h}_{125} \rightarrow \text{a}\text{a} \rightarrow \mu^{-}\mu^{+} c\bar{c})$, for pseudoscalar masses between 4 and 11\GeV. The limits are computed using the modified frequentist $\text{CL}_{\text{s}}$ approach~\cite{Junk:1999kv,A:L:Read_2002} employing an asymptotic approximation to the distribution of the profile likelihood ratio test statistic~\cite{CMS:2018mlc}. The results are presented in figure~\ref{fig:Limits-BR}, and they corroborate the above observation made when performing a maximum-likelihood fit under the background-only hypothesis, namely, that no significant excesses are seen. Only a few minor and local deviations, very close to the two-standard-deviation threshold are preferred by the data when scanning for potential signal hypotheses. As a result, the median expected upper limit ranges from $5.0 \times 10^{-4}$ for a mass close to \ma = 4\GeV up to a value of $1.4 \times 10^{-3}$ for a mass hypothesis in the vicinity of \ma = 9.5\GeV, while the observed upper limits ranges from $3.3 \times 10^{-4}$ for a mass close to \ma = 4.8\GeV up to a value of $3.1 \times 10^{-3}$ for \ma = 9.5\GeV.

As can be noted, the analysis expected sensitivity slightly degrades as the mass of the light boson increases. This can be explained by the fact that, despite the background is falling roughly exponentially from $\ma \approx 4\GeV$ to $\ma  \approx 9\GeV$, the narrower signal peak featured by lighter states, makes the overall discrimination more prominent for masses in the lower range of the search interval. From $\ma  \approx 9\GeV$ to $\ma  \approx 10.5\GeV$, the resonant components of the background predominate over the exponential continuum, which consequently engenders a further deterioration of the expected limits in that mass range.

\begin{figure}[htp]
    \centering
     \centering
    \begin{subfigure}{0.75\textwidth}
        \centering
        \includegraphics[width=0.95\textwidth]{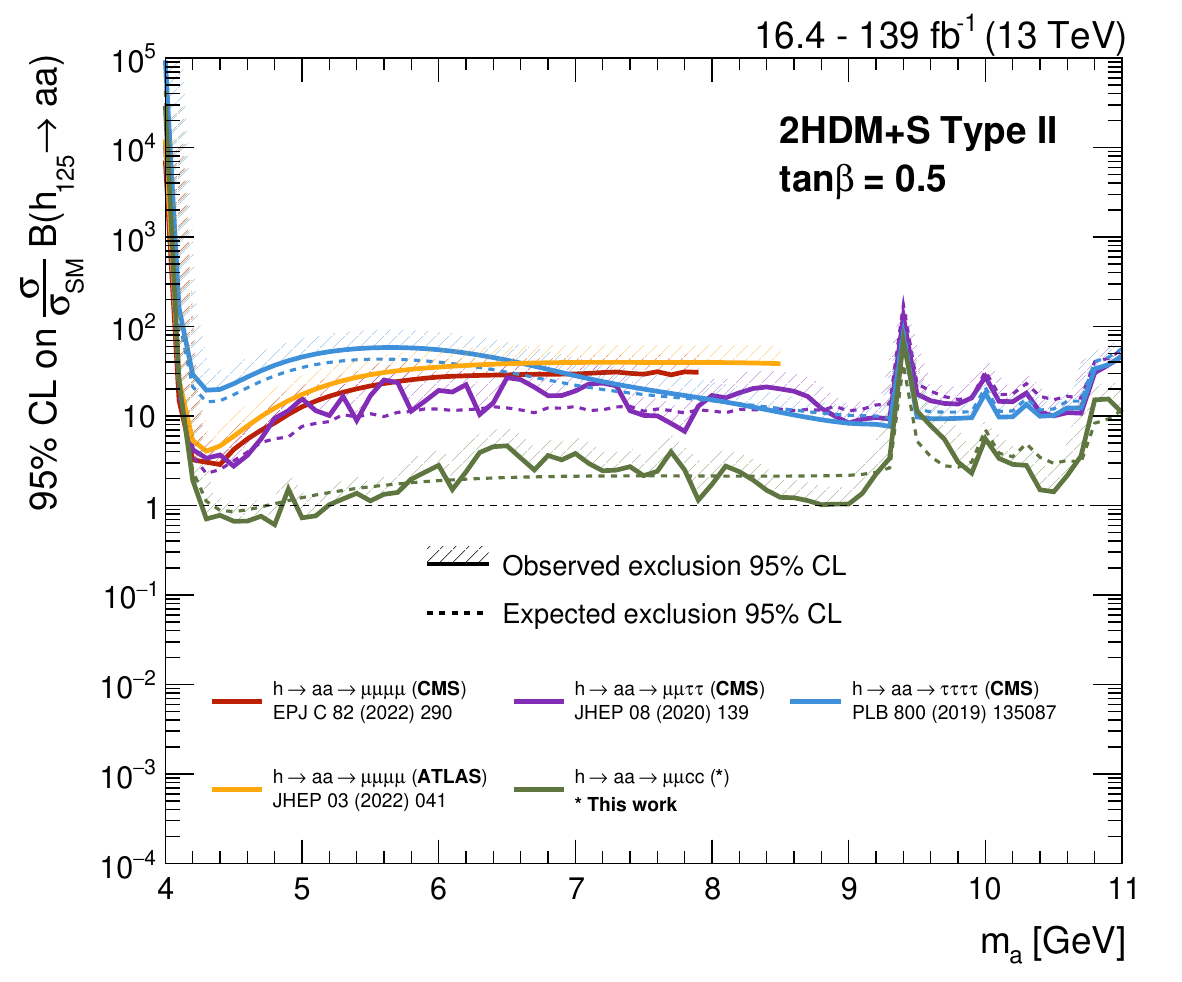}
    \end{subfigure}
    \begin{subfigure}{0.75\textwidth}
        \centering
        \includegraphics[width=0.95\textwidth]{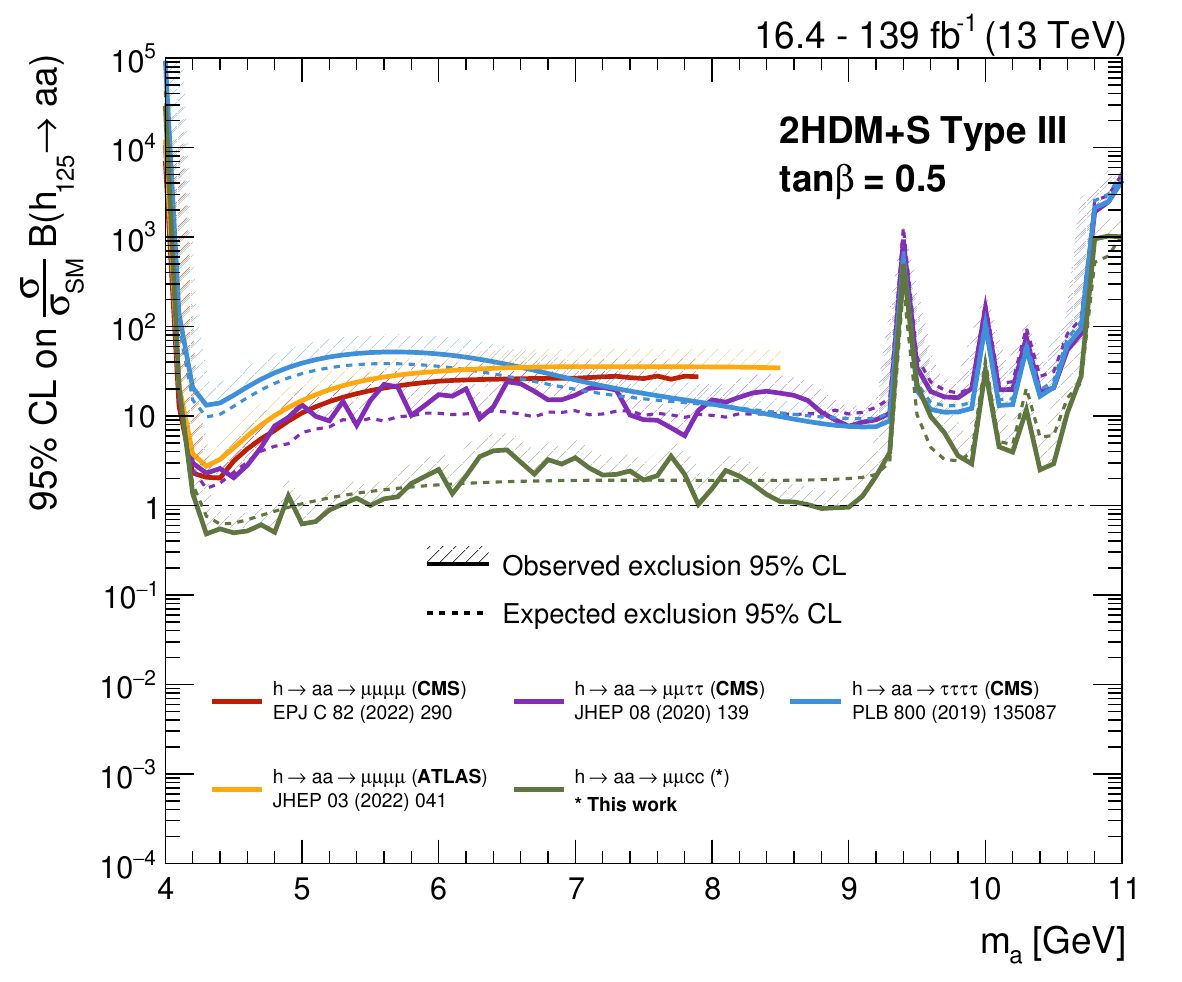}
    \end{subfigure}
    \caption{Observed and expected upper limits at 95\%~CL on $\sigma/\sigma_{\text{SM}} \cdot \mathcal{B}(\text{h}_{125} \rightarrow \text{a}\text{a})$ as a function of \ma for Type II (upper) and Type III (lower) 2HDM+S scenarios. The limits are computed assuming a value of $\tan\beta = 0.5$. The results of this search employing the CMS Open Data (16.4~\invfb) are compared to several experimental results delivered by both the CMS and the ATLAS collaborations in the $\mu\mu\mu\mu$, $\mu\mu\tau\tau$, and $\tau\tau\tau\tau$ final states. The CMS and ATLAS results comprise at least 35.9~\invfb of data. }
    \label{fig:2HDMS-exclusions}
\end{figure}

The above limits presented in figure~\ref{fig:Limits-BR} can be regarded as model-independent under the assumption that the narrow width approximation is valid for all resonances involved in the decay chain -- a condition that is comfortably applicable to both $h_{125}$ and $\text{a}$. The results are thus translated into model-dependent constraints on $\sigma/\sigma_{\text{SM}} \cdot \mathcal{B}(\text{h}_{125} \rightarrow \text{a}\text{a})$ as a function of \ma, and assuming a value of $\tan\beta = 0.5$, for Type II and III scenarios of the 2HDM+S. The choice for this particular value of $\tan\beta$ is motivated by the observations made in section~\ref{sec:intro}, and it exemplifies a point embedded in the phase-space region for which both model types above feature a preferred coupling of the light pseudoscalar to up-type quarks, or simultaneously, to both up-type and down-type quarks. For the reinterpretation of the results shown in figure~\ref{fig:Limits-BR}, the model branching fractions for pseudoscalar decays to a pair of fermions, $\mathcal{B}(\text{a} \rightarrow \text{f}\bar{\text{f}})$, were taken from~\cite{Haisch:2018kqx}. To profit from a higher mass granularity compared to the existing scanned mass points, a spline interpolation has been performed for both the theoretical predictions of $\mathcal{B}(\text{a} \rightarrow \text{f}\bar{\text{f}})$ and the available experimental limits. The upper limits at 95\%~CL on $\sigma/\sigma_{\text{SM}} \cdot \mathcal{B}(\text{h}_{125} \rightarrow \text{a}\text{a})$ can be found in figure~\ref{fig:2HDMS-exclusions}, along with a comparison with the sensitivity of several experimental results mentioned in section~\ref{sec:intro}, and that are relevant for the mass range probed in this work. Specifically, results from both the CMS and the ATLAS collaborations using at least 35.9~\invfb of data that cover the $\mu\mu\mu\mu$, $\mu\mu\tau\tau$, and $\tau\tau\tau\tau$ final states are included in the figure along with the results obtained in this analysis.

As can be observed in figure~\ref{fig:2HDMS-exclusions}, for these two 2HDM+S scenarios, the constraints imposed by this analysis are much more stringent than those of the other existing analyses for the same mass range. In fact, it can be said that this analysis offers the first physical constraints for $\mathcal{B}(\text{h}_{125} \rightarrow \text{a}\text{a})$, if a similar cross section to the one predicted in the SM for the Higgs is assumed. The stronger sensitivity of the $\mu\mu cc$ channel compared to fully leptonic final states can be understood if one considers that for the 2HDM+S configurations shown in figure~\ref{fig:2HDMS-exclusions}, the light boson predominantly decays to a pair of c-quarks, given its preferred coupling to up-type quarks. Leptonic decays of the pseudoscalar, despite the experimental advantages offered, are highly suppressed in those particular 2HDM+S scenarios. A combination of high decay rates via the $\text{a} \rightarrow c\bar{c}$ mode and the excellent experimental reconstruction achieved when probing $\text{a} \rightarrow \mu^{-}\mu^{+}$ decays thus delivers a more powerful constraint in the above-depicted 2HDM+S parameter phase space. Up to the present moment, this search is the only experimental result that is able to cross the unity threshold for $\mathcal{B}(\text{h}_{125} \rightarrow \text{a}\text{a})$, under the specific model configurations explained above. These constraints are still limited to a small mass interval though. However, it is still plausible to think of further improvements for the mass range corresponding to heavier pseudoscalars, as more of the data that has already been recorded in Run 2 and during Run 3 could be added. The development of dedicated c-taggers optimized for the specific topology studied here could also contribute substantially to strengthening the potential of such a challenging but fascinating final state.


\section{Summary}
\label{sec:summary}


The first search for exotic decays of the 125 GeV Higgs boson ($\text{h}_{125}$) into a pair of light bosons (\text{a}) in the ${\text{h}_{125} \rightarrow \text{a}\text{a} \rightarrow \mu^{-}\mu^{+} c\bar{c}}$ channel has been presented. A publicly available dataset of proton-proton collisions collected in 2016 by the CMS experiment at a center-of-mass energy of 13\TeV, corresponding to a total integrated luminosity of 16.4~\invfb, was analyzed.

The analysis exploits, for the first time in these types of searches, the prospect of charm jet identification techniques when applied to collimated and low-mass ${\text{a} \rightarrow c\bar{c}}$ systems. The current c-tagging methods employed by the CMS collaboration, even when not mainly designed to tackle such a class of topologies, can identify a variety of these scenarios with adequate efficiency. The above, when combined with the powerful di-muon mass resolution exhibited by the CMS detector, allows us to reach considerable levels of sensitivity for this kind of process.

No sign of decays of the 125 GeV scalar into a pair of pseudoscalars via the channel investigated here has been observed. The results are thus presented in terms of 95\%~CL upper limits on the product of the cross section and branching fraction relative to the standard model (SM) Higgs boson production cross section, i.e. ${\sigma/\sigma_{\text{SM}} \cdot \mathcal{B}(\text{h}_{125} \rightarrow \text{a}\text{a} \rightarrow \mu^{-}\mu^{+} c\bar{c})}$. The above limits are translated into model-specific constraints on ${\sigma/\sigma_{\text{SM}} \cdot \mathcal{B}(\text{h}_{125} \rightarrow \text{a}\text{a} )}$ in the context of Type II and III two Higgs doublets plus singlet models (2HDM+S). The exclusion limits established by this search are compared to several experimental results obtained by the ATLAS and CMS collaborations in other decay channels, which make use of sizably larger datasets. By probing those 2HDM+S configurations where the coupling of up-type quarks to the light pseudoscalar is enhanced, it is demonstrated that this search produces the most stringent constraints up to date for those model realizations.


\section*{Acknowledgements}
\sloppy{
D.P.A. would like to thank his colleagues from the CMS collaboration for releasing about half of the total dataset collected in 2016, as well as a large amount of different simulated processes as part of the Open Data. D.P.A. acknowledges the dedicated effort to document the details of the usage of the software, detector calibration, and other relevant aspects of the released data through the various public web portals. Furthermore, D.P.A. would like to express his gratitude to the CMS collaboration for permitting him to pursue this publication under individual authorship. This work is possible thanks to the outstanding performance of the LHC through these years and the impressive computing infrastructure deployed and supported by CERN. This work has been supported by the Bundesministerium f{\"u}r Bildung und Forschung. Finally, D.P.A. genuinely values the discussions with his colleagues from the I. Physikalisches Institut B at RWTH Aachen and the insightful comments received for the completion of this manuscript.
}


\clearpage
\printbibliography

\end{document}